# A 1D Radiative Transfer Benchmark with Polarization via Doubling and Adding

B. D. Ganapol
Department of Aerospace and Mechanical Engineering
University of Arizona

**ABSTRACT**
Highly precise numerical solutions to the radiative transfer equation with polarization present a special challenge. Here, we establish a precise numerical solution to the radiative transfer equation with combined Rayleigh and isotropic scattering in a 1D-slab medium with simple polarization. The 2- Stokes vector solution for the fully discretized radiative transfer equation in space and direction derives from the method of doubling and adding enhanced through convergence acceleration. Updates to benchmark solutions found in the literature to 7 places for reflectance and transmittance as well as for angular flux follow. Finally, we conclude with the numerical solution in a partially randomly absorbing heterogeneous medium.

**INTRODUCTION**
Establishing precise benchmarks for radiative transfer with polarization remains a challenge, even in today's advanced computational environment. With polarization, the radiative transfer equation takes on added mathematical richness and therefore added challenge. In particular, the radiative transfer equation, becomes a vector equation for the Stokes or surrogate Stokes-vectors describing oscillations of the electric and magnetic fields underpinning light propagation. References 1 to 6 exemplify the numerous investigations concerning the solution of the radiative transfer equation with polarization. What is to be presented here is yet another one. While considering combined Rayleigh and isotropic scattering only for the two Stokes vector, establishment of precision benchmarks attempts to compensate for the apparent simplicity. The numerical solution for the fully discretized radiative transfer equation originates from the method of doubling and adding coupled with convergence acceleration both in spatial and directional (angular) discretizations. The method of doubling and adding has been applied to polarization previously [7,8], but most likely never to the precise extent found here. The advantage of the doubling approach is its simplicity. No eigen- values or vectors required and the scattering matrix remains as an exact kernel avoiding approximate Legendre polynomial expansions. In addition, through a discrete delta function representation, no particular solutions are necessary as is common for most existing numerical transport methods. Finally, we show the conservative case is on equal footing with the non-conservative, which is the not necessarily true for all numerical transport solutions found in the literature.

The presentation proceeds as follows. The analysis begins with the radiative transfer equation for the two Stokes vector (generalized) intensity for combined Rayleigh and isotropic scattering. Directional discretization and continuous spatial variation lead to a set of coupled ODEs for the vector intensity. The ODEs, solved over a spatial node of uniform properties, gives a response matrix relating the incoming and outgoing intensities. Next, the symmetry of the nodal response is derived before combining the single node response with a general composite slab response. The combined nodal response provides the basis for doubling responses to determine the overall homogeneous medium response. This, in turn, gives the exiting intensities of a polarizing medium bordering a Lambertian reflector with beam illumination. In addition, angular edits in the interior result from knowledge of the overall response through backward recurrence. With numerical implementation focused on precision by considering various single node responses algorithms, beam source implementation, the grazing direction and convergence acceleration, we develop a precise numerical transport solution. We apply the solution to update two benchmarks from the literature, present a conservative benchmark and one for a partially randomly absorbing medium.

## I. Theory
### A. Transport Equation with Polarization

We begin with the azimuthally integrated 1D transport equation with polarization in plane slab geometry for the vector intensity $I(\tau,\mu)$ at position $\tau$ measured in optical depths and in direction $\mu$ (cosine of the polar angle in direction $\tau$)

$$\mu\frac{\partial}{\partial\tau}I(\tau,\mu)+I(\tau,\mu)=\frac{1}{2}\omega Q(\mu)\int_{-1}^{1}d\mu'Q^{T}(\mu')I(\tau,\mu');\ \tau\in[0,\tau_0], \qquad (1a)$$

where we assume a coupled Rayleigh/isotropically scattering model [9-11]. At this point, we only note scattering symmetry in incoming and outgoing directions $\mu$,

$$Q(-\mu)=Q(\mu), \qquad (1b)$$

where $Q(\mu)$, explicitly expressed below, is the scattering matrix factor. The 2-Stokes vector intensity distribution at any position $\tau$ within a homogeneous slab,

$$\boldsymbol{I}(\tau,\mu) \equiv \begin{bmatrix} I(\tau,\mu) \\ Q(\tau,\mu) \end{bmatrix}, \quad (1c)$$

arises from an incoming beam in direction $\mu_0$ entering the near surface,

$$\boldsymbol{I}(0,\mu) = \frac{1}{2}\begin{bmatrix} F_I \\ F_Q \end{bmatrix}\delta(\mu-\mu_0); \ \mu \geq 0; \quad (1d)$$

and Lambertian reflection on the far surface

$$\boldsymbol{I}(\tau_0,-\mu) = 2\lambda_0 \boldsymbol{D}\int_0^1 d\mu'\mu' \boldsymbol{I}(\tau_0,\mu'); \ \mu \geq 0. \quad (1e)$$

$\lambda_0$ is the reflection coefficient and $\boldsymbol{D}$ is the reflective coupling matrix between the two components. Finally, $\omega$ is the single scatter albedo of a homogeneous slab of thickness $\tau_0$.

## B. Discrete Ordinance Balance Equation

One finds the angularly discretized form of Eqs(1) by introducing $2N$ angular discretizations ($\mu_m > 0$)

$$\mu = \begin{cases} \mu_m, \ m=1,...,N \\ -\mu_m, \ m=N+1,...,2N \end{cases} \quad (2a)$$

in forward (+) and backward (−) directions. The quadrature abscissae come from the usual half-range Gauss quadrature approximation. In addition, assuming the half-range quadrature for angular integration over $N$ forward and $N$ backward directions gives the discrete balance in a direction $\mu_m$

$$\mu_m \frac{d}{d\tau}\boldsymbol{I}_m(\tau) + \boldsymbol{I}_m(\tau) = \frac{1}{2}\omega \sum_{m'=1}^{2N} \omega_{m'}\boldsymbol{Q}(\mu_m)\boldsymbol{Q}^T(\mu_{m'})\boldsymbol{I}_{m'}(\tau). \quad (2b)$$

These equations are the discrete ordinates (balance) equations. The theory of Lagrange interpolation enables an estimation of the discretization error as a contour integration and confirms convergence in *N* for sufficiently smooth integrands, which scattering guarantees.

If the directed intensity vectors in the forward and backward directions are

$$\boldsymbol{I}^{+}(\tau) \equiv \begin{bmatrix} \boldsymbol{I}_{1}^{T}(\tau) & \boldsymbol{I}_{2}^{T}(\tau) & \ldots & \boldsymbol{I}_{N}^{T}(\tau) \end{bmatrix}^{T}$$
$$\boldsymbol{I}^{-}(\tau) \equiv \begin{bmatrix} \boldsymbol{I}_{N+1}^{T}(\tau) & \boldsymbol{I}_{N+2}^{T}(\tau) & \ldots & \boldsymbol{I}_{2N}^{T}(\tau) \end{bmatrix}^{T},$$
(3a,b)

then

$$-\frac{d}{d\tau}\boldsymbol{I}^{-}(\tau) + \boldsymbol{M}^{-1}(\boldsymbol{I} - \boldsymbol{PW})\boldsymbol{I}^{-}(\tau) = \boldsymbol{M}^{-1}\boldsymbol{PW}\boldsymbol{I}^{+}(\tau)$$
$$\frac{d}{d\tau}\boldsymbol{I}^{+}(\tau) + \boldsymbol{M}^{-1}(\boldsymbol{I} - \boldsymbol{PW})\boldsymbol{I}^{+}(\tau) = \boldsymbol{M}^{-1}\boldsymbol{PW}\boldsymbol{I}^{-}(\tau).$$
(3c,d)

The vector equation becomes equivalent to the scalar formulation (without polarization) [12] by partitioning the matrices in Eqs(3c,d) to account for the two-component nature of the intensity in each direction as follows:

$$\boldsymbol{W} \equiv diag\left\{ \begin{bmatrix} \omega_{m} & 0 \\ 0 & \omega_{m} \end{bmatrix}; m = 1,\ldots,N \right\}$$
$$\boldsymbol{M} \equiv diag\left\{ \begin{bmatrix} \mu_{m} & 0 \\ 0 & \mu_{m} \end{bmatrix}; m = 1,\ldots,N \right\}$$
(3e,f)

and a matrix of 2 by 2 blocks

$$\boldsymbol{P} \equiv \frac{\omega}{2}\left\{ \boldsymbol{Q}(\mu_{m})\boldsymbol{Q}^{T}(\mu_{m'}); m,m' = 1,\ldots,N \right\}.$$
(3g)

This effectively unrolls the 2-vector and Eqs(3c,d) become

$$\frac{d}{d\tau}\boldsymbol{I}^{\pm}(\tau) = \mp\boldsymbol{M}^{-1}(\boldsymbol{I} - \boldsymbol{PW})\boldsymbol{I}^{\pm}(\tau) \pm \boldsymbol{M}^{-1}\boldsymbol{PW}\boldsymbol{I}^{\mp}(\tau),$$
(4±)

where we have replaced $N$ by $2N$ and $\boldsymbol{I}$ is the conformable identity matrix of order $N$. In this way, the vector characterization of polarization directly folds into the discretization by including an angular Stokes vector of stride two.

When the outgoing and incoming vectors are stacked

$$\boldsymbol{Y}(\tau) \equiv \left[ \boldsymbol{I}^{T+}(\tau) \quad \boldsymbol{I}^{T-}(\tau) \right]^T, \tag{5a}$$

Eqs(4±) conveniently become

$$\frac{d}{d\tau}\boldsymbol{Y}(\tau) + \boldsymbol{A}\boldsymbol{Y}(\tau) = 0 \tag{5b}$$

with constant Jacobian

$$\boldsymbol{A} = \begin{bmatrix} \boldsymbol{M}^{-1}(\boldsymbol{I} - \boldsymbol{PW}) & -\boldsymbol{M}^{-1}\boldsymbol{PW} \\ \boldsymbol{M}^{-1}\boldsymbol{PW} & -\boldsymbol{M}^{-1}(\boldsymbol{I} - \boldsymbol{PW}) \end{bmatrix}. \tag{5c}$$

We next solve Eqs(5) over a single homogeneous node of width $h$.

## C. Single Node Response

Because the 2-vector has been unrolled, the entire discretization scheme of the scalar treatment is available. In the current context, a node, shown in Fig. 1, is a computational unit within the slab medium. The output response of the single node $\boldsymbol{I}^-(\tau_j), \boldsymbol{I}^+(\tau_{j+1})$ from each interface from input $\boldsymbol{I}^+(\tau_j), \boldsymbol{I}^-(\tau_{j+1})$ into each interface and will come from the discrete ordinates equations, Eqs(5).

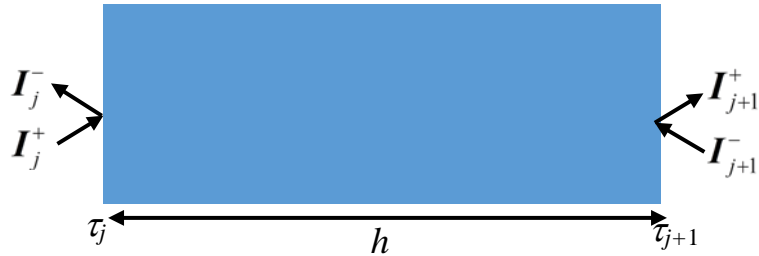

Fig. 1. Fundamental node of width $h$.

Since the node is homogeneous, $A$ is constant and the analytical solution of Eq(5b) over the node $[\tau_j, \tau_{j+1}]$ is simply

$$Y_{j+1} = e^{-Ah}Y_j, \tag{6a}$$

where

$$Y_j \equiv \begin{bmatrix} I_j^{T+} & I_j^{T-} \end{bmatrix}^T \tag{6b}$$

and $e^{-Ah}$ is the matrix exponential. We now find the connection between what enters and exits the node through a rearrangement of vector components in Eq(6a).

If the matrix exponential is partitioned

$$e^{-Ah} = \begin{bmatrix} E_{11}(h) & E_{12}(h) \\ E_{21}(h) & E_{22}(h) \end{bmatrix}, \tag{7}$$

then Eq(6a) is

$$\begin{bmatrix} I_{j+1}^+ \\ I_{j+1}^- \end{bmatrix} = \begin{bmatrix} E_{11} & E_{12} \\ E_{21} & E_{22} \end{bmatrix} \begin{bmatrix} I_j^+ \\ I_j^- \end{bmatrix}, \tag{8}$$

where the $h$ dependence of the partitions has been suppressed. By identifying the outgoing and ingoing vectors at each surface as

$$\begin{bmatrix} I_j^- \\ I_{j+1}^+ \end{bmatrix}, \begin{bmatrix} I_{j+1}^- \\ I_j^+ \end{bmatrix}, \tag{9}$$

Eq(6a) becomes

$$\begin{bmatrix} E_{12} & -I \\ E_{22} & 0 \end{bmatrix} \begin{bmatrix} I_j^- \\ I_{j+1}^+ \end{bmatrix} = \begin{bmatrix} 0 & -E_{11} \\ I & -E_{21} \end{bmatrix} \begin{bmatrix} I_{j+1}^- \\ I_j^+ \end{bmatrix}. \tag{10}$$

Finally, assuming suitable inevitability,

$$\begin{bmatrix} I_j^- \\ I_{j+1}^+ \end{bmatrix} = R \begin{bmatrix} I_{j+1}^- \\ I_j^+ \end{bmatrix}, \tag{11a}$$

the response matrix emerges as

$$R \equiv \begin{bmatrix} E_{12} & -I \\ E_{22} & 0 \end{bmatrix}^{-1} \begin{bmatrix} 0 & -E_{11} \\ I & -E_{21} \end{bmatrix}. \tag{11b}$$

An interpretation of the response matrix is now in order.

**1. Collision symmetry: Physical argument**
If $R$ is partitioned as

$$R \equiv \begin{bmatrix} R_{11} & R_{12} \\ R_{21} & R_{22} \end{bmatrix}, \tag{12a}$$

then Eq(11a) expanded is

$$\begin{aligned} I_j^- &= R_{11} I_{j+1}^- + R_{12} I_j^+ \\ I_{j+1}^+ &= R_{21} I_{j+1}^- + R_{22} I_j^+. \end{aligned} \tag{12b}$$

The physical interpretation of Eq(12b) says the exiting intensity from surface $j$ $(I_j^-)$ is a combination of the transmitted intensity from surface $j+1$ $(I_{j+1}^-)$ and the reflected intensity from the intensity entering surface $j$ $(I_j^+)$. Hence, $R_{11}$ and $R_{12}$ are transmittance and reflectance partitions respectively. Similarly, the transmitted intensity from surface $j+1$ $(I_{j+1}^+)$ comes from reflection of the intensity entering at surface $j+1$ $\left(\boldsymbol{I}_{j+1}^{-}\right)$ and that transmitted from surface $j$ $\left(\boldsymbol{I}_{j}^{+}\right)$. Therefore, $\boldsymbol{R}_{21}$ and $\boldsymbol{R}_{22}$ are reflectance and transmittance partitions giving the exiting intensity $\boldsymbol{I}_{j+1}^{+}$. Since the medium is homogeneous, the reflectance and transmittance partitions must be identical

$$\begin{aligned}\boldsymbol{R}_{f} &\equiv \boldsymbol{R}_{21} = \boldsymbol{R}_{12}\\ \boldsymbol{T}_{n} &\equiv \boldsymbol{R}_{11} = \boldsymbol{R}_{22}.\end{aligned} \tag{13}$$

Thus, the response matrix is symmetric

$$\boldsymbol{R} \equiv \begin{bmatrix} \boldsymbol{T}_{n} & \boldsymbol{R}_{f} \\ \boldsymbol{R}_{f} & \boldsymbol{T}_{n} \end{bmatrix}. \tag{14}$$

Can this be shown mathematically?

## 2. Collision symmetry: Mathematical derivation
From the definition of $\boldsymbol{R}$ in Eq(12a)

$$\boldsymbol{R} \equiv \boldsymbol{T}^{-1} \begin{bmatrix} \boldsymbol{0} & -\boldsymbol{E}_{11} \\ \boldsymbol{I} & -\boldsymbol{E}_{21} \end{bmatrix} \tag{15a}$$

with

$$\boldsymbol{T} \equiv \begin{bmatrix} \boldsymbol{E}_{12} & -\boldsymbol{I} \\ \boldsymbol{E}_{22} & \boldsymbol{0} \end{bmatrix}. \tag{15b}$$

Therefore

$$\begin{bmatrix} \boldsymbol{E}_{12} & -\boldsymbol{I} \\ \boldsymbol{E}_{22} & \boldsymbol{0} \end{bmatrix} \boldsymbol{T}^{-1} = \begin{bmatrix} \boldsymbol{I} & \boldsymbol{0} \\ \boldsymbol{0} & \boldsymbol{I} \end{bmatrix}. \tag{15c}$$

With some algebra, one finds

$$T^{-1} = \begin{bmatrix} 0 & E_{22}^{-1} \\ -I & E_{12}E_{22}^{-1} \end{bmatrix}; \qquad (16a)$$

and therefore

$$R \equiv \begin{bmatrix} 0 & E_{22}^{-1} \\ -I & E_{12}E_{22}^{-1} \end{bmatrix} \begin{bmatrix} 0 & -E_{11} \\ I & -E_{21} \end{bmatrix} = \begin{bmatrix} E_{22}^{-1} & -E_{22}^{-1}E_{21} \\ E_{12}E_{22}^{-1} & E_{11} - E_{12}E_{22}^{-1}E_{21} \end{bmatrix}. \qquad (16b)$$

To recognize the desired symmetry, the relation of the partitions to matrix $A$ requires closer examination.

One can express the matrix exponential in a variety of ways as will be shown below. However, to gain theoretical insight into the response matrix, it is sufficient to apply the formal Taylor series definition

$$e^{-Ah} \equiv I + \sum_{j=1}^{\infty} \frac{(-1)^j h^j}{j!} A^j = \begin{bmatrix} E_{11}(h) & E_{12}(h) \\ E_{21}(h) & E_{22}(h) \end{bmatrix}. \qquad (17)$$

Since

$$A = \begin{bmatrix} \alpha & -\beta \\ \beta & -\alpha \end{bmatrix} \equiv \begin{bmatrix} \eta_{11}^{(1)} & -\eta_{12}^{(1)} \\ \eta_{12}^{(1)} & -\eta_{11}^{(1)} \end{bmatrix}, \qquad (18a)$$

the squared matrix is symmetric

$$A^2 = \begin{bmatrix} \alpha & -\beta \\ \beta & -\alpha \end{bmatrix} \begin{bmatrix} \alpha & -\beta \\ \beta & -\alpha \end{bmatrix} = \begin{bmatrix} \alpha^2 - \beta^2 & -\alpha\beta + \beta\alpha \\ \beta\alpha - \alpha\beta & -\beta^2 + \alpha^2 \end{bmatrix} = \begin{bmatrix} \gamma_{11}^{(1)} & \gamma_{12}^{(1)} \\ \gamma_{12}^{(1)} & \gamma_{11}^{(1)} \end{bmatrix} \qquad (18b)$$

and, as shown in Appendix A, leads to the symmetry of even powers

$$A^{2m} = \begin{bmatrix} \gamma_{11}^{(m)} & \gamma_{12}^{(m)} \\ \gamma_{12}^{(m)} & \gamma_{11}^{(m)} \end{bmatrix}. \qquad (18c)$$

Similarly, for odd powers the pattern of positive and negative elements is invariant

$$A^{2m-1} = \begin{bmatrix} \eta_{11}^{(m)} & -\eta_{12}^{(m)} \\ \eta_{12}^{(m)} & -\eta_{11}^{(m)} \end{bmatrix}.$$

When introduced into Eq(17), there results

$$E_{11} \equiv I + \sum_{m=1}^{\infty} \left[ \frac{h^{2m}}{(2m)!} \gamma_{11}^{(m)} + \frac{h^{2m-1}}{(2m-1)!} \eta_{11}^{(m)} \right]$$

$$E_{12} \equiv \sum_{m=1}^{\infty} \left[ \frac{h^{2m}}{(2m)!} \gamma_{12}^{(m)} - \frac{h^{2m-1}}{(2m-1)!} \eta_{12}^{(m)} \right]$$

$$E_{21} \equiv \sum_{m=1}^{\infty} \left[ \frac{h^{2m}}{(2m)!} \gamma_{12}^{(m)} + \frac{h^{2m-1}}{(2m-1)!} \eta_{12}^{(m)} \right]$$

$$E_{22} \equiv I + \sum_{m=1}^{\infty} \left[ \frac{h^{2m}}{(2m)!} \gamma_{11}^{(m)} - \frac{h^{2m-1}}{(2m-1)!} \eta_{11}^{(m)} \right].$$

(19a,b,c,d)

Note again we have suppressed the $h$ dependence.

To continue, there is an alternative representation of Eq(6a), which reads

$$e^{Ah} Y_{j+1} = Y_j, \tag{20}$$

and from Eqs(7) and (19)

$$e^{Ah} = \begin{bmatrix} E_{11}(-h) & E_{12}(-h) \\ E_{21}(-h) & E_{22}(-h) \end{bmatrix} = \begin{bmatrix} E_{22} & E_{21} \\ E_{12} & E_{11} \end{bmatrix}, \tag{21a}$$

which on rearrangement of Eq(20) gives

$$\begin{bmatrix} \mathbf{0} & \mathbf{E}_{22} \\ -\mathbf{I} & \mathbf{E}_{12} \end{bmatrix} \begin{bmatrix} \mathbf{I}_j^- \\ \mathbf{I}_{j+1}^+ \end{bmatrix} = \begin{bmatrix} -\mathbf{E}_{21} & \mathbf{I} \\ -\mathbf{E}_{11} & \mathbf{0} \end{bmatrix} \begin{bmatrix} \mathbf{I}_{j+1}^- \\ \mathbf{I}_j^+ \end{bmatrix}. \tag{21b}$$

Therefore, with some algebra,

$$\begin{bmatrix} \mathbf{I}_j^- \\ \mathbf{I}_{j+1}^+ \end{bmatrix} = \mathbf{R} \begin{bmatrix} \mathbf{I}_{j+1}^- \\ \mathbf{I}_j^+ \end{bmatrix}, \tag{22a}$$

and

$$\mathbf{R} = \begin{bmatrix} \mathbf{0} & \mathbf{E}_{22} \\ -\mathbf{I} & \mathbf{E}_{12} \end{bmatrix}^{-1} \begin{bmatrix} -\mathbf{E}_{21} & \mathbf{I} \\ -\mathbf{E}_{11} & \mathbf{0} \end{bmatrix}. \tag{22b}$$

Then, by finding the inverse explicitly, as above

$$\mathbf{R} = \begin{bmatrix} \mathbf{E}_{11} - \mathbf{E}_{12}\mathbf{E}_{22}^{-1}\mathbf{E}_{21} & \mathbf{E}_{12}\mathbf{E}_{22}^{-1} \\ -\mathbf{E}_{22}^{-1}\mathbf{E}_{21} & \mathbf{E}_{22}^{-1} \end{bmatrix}. \tag{23}$$

On comparison of the two $\mathbf{R}$s of Eqs(23) and (16b), the desired symmetry emerges

$$\begin{aligned} \mathbf{T}_n &\equiv \mathbf{E}_{22}^{-1} = \mathbf{E}_{11} - \mathbf{E}_{12}\mathbf{E}_{22}^{-1}\mathbf{E}_{21} \\ \mathbf{R}_f &\equiv -\mathbf{E}_{22}^{-1}\mathbf{E}_{21} = \mathbf{E}_{12}\mathbf{E}_{22}^{-1}. \end{aligned} \tag{24}$$

As now shown, the single node response enables a combined node response.

**D. Combined Node Response Matrix**

Figure 2 shows the addition of a node of width $h$ to $l$-1 composite nodes for which one knows a collective response matrix. Essentially, one follows the invariant embedding approach first suggested by V. Ambartzmanian and subsequently popularized by Chandrasekhar.

The composite of $l$-1 nodes of arbitrary widths assumes a combined response matrix of $\mathbf{Q}_{l-1}$ to give

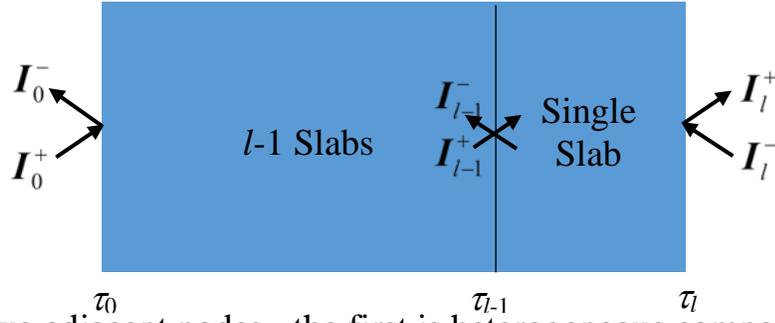

Fig. 2. Two adjacent nodes– the first is heterogeneous composed of $l-1$ nodes and the second is a single homogeneous node.

$$\begin{bmatrix} I_0^- \\ I_{l-1}^+ \end{bmatrix} = Q_{l-1} \begin{bmatrix} I_{l-1}^- \\ I_0^+ \end{bmatrix}. \tag{25a}$$

Taking advantage of response matrix symmetry, gives the following order $N$ partitioned matrix response:

$$Q_{l-1} = \begin{bmatrix} Q_{l-1,1} & Q_{l-1,2} \\ Q_{l-1,2} & Q_{l-1,1} \end{bmatrix}., \tag{25b}$$

The composite can have heterogeneous nodes with properties differing from, or the same as the added homogeneous node. For the added node

$$\begin{bmatrix} I_{l-1}^- \\ I_l^+ \end{bmatrix} = R \begin{bmatrix} I_l^- \\ I_{l-1}^+ \end{bmatrix}. \tag{26a}$$

Expanded, Eqs(26a) are

$$\begin{aligned} I_0^- &= Q_{l-1,1} I_{l-1}^- + Q_{l-1,2} I_0^+ \\ I_l^+ &= R_f I_l^- + T_n I_{l-1}^+ \\ I_{l-1}^+ &= Q_{l-1,2} I_{l-1}^- + Q_{l-1,1} I_0^+ \\ I_{l-1}^- &= T_n I_l^- + R_f I_{l-1}^+. \end{aligned} \tag{26b}$$

On rearrangement of the last two and first two equations respectively, there results

$$\begin{bmatrix} I & -Q_{l-1,2} \\ -R_f & I \end{bmatrix} \begin{bmatrix} I_{l-1}^+ \\ I_{l-1}^- \end{bmatrix} = \begin{bmatrix} 0 & -Q_{l-1,1} \\ -T_n & 0 \end{bmatrix} \begin{bmatrix} I_l^- \\ I_0^+ \end{bmatrix}, \tag{27a}$$

$$\begin{bmatrix} I_0^- \\ I_l^+ \end{bmatrix} = \begin{bmatrix} 0 & Q_{l-1,2} \\ R_f & 0 \end{bmatrix} \begin{bmatrix} I_l^- \\ I_0^+ \end{bmatrix} + \begin{bmatrix} 0 & Q_{l-1,1} \\ T_n & 0 \end{bmatrix} \begin{bmatrix} I_{l-1}^+ \\ I_{l-1}^- \end{bmatrix}. \tag{27b}$$

Solving for the leftmost vector in the last equation gives

$$\begin{bmatrix} I_{l-1}^+ \\ I_{l-1}^- \end{bmatrix} = U_l \begin{bmatrix} I_l^- \\ I_0^+ \end{bmatrix}, \tag{28a}$$

with

$$U_l \equiv w_{m,l}^{-1} w_{p,l} \tag{28b}$$

$$w_{m,l} \equiv \begin{bmatrix} I & -Q_{l-1,2} \\ -R_f & I \end{bmatrix}$$

$$w_{p,l} \equiv \begin{bmatrix} 0 & Q_{l-1,1} \\ T_n & 0 \end{bmatrix}. \tag{28c}$$

When Eq(28a) is introduced into Eq(27b),

$$\begin{bmatrix} I_0^- \\ I_l^+ \end{bmatrix} = \left\{ \begin{bmatrix} 0 & Q_{l-1,1} \\ T_n & 0 \end{bmatrix} U_l + \begin{bmatrix} 0 & Q_{l-1,2} \\ R_f & 0 \end{bmatrix} \right\} \begin{bmatrix} I_l^- \\ I_0^+ \end{bmatrix},$$

from which the combined response matrix (a recurrence of partitioned matrices) now for $l$ nodes follows

$$Q_l = \begin{bmatrix} 0 & Q_{l-1,1} \\ T_n & 0 \end{bmatrix} U_l + \begin{bmatrix} 0 & Q_{l-1,2} \\ R_f & 0 \end{bmatrix}, \tag{28d}$$

and therefore

$$\begin{bmatrix} I_0^- \\ I_l^+ \end{bmatrix} = Q_l \begin{bmatrix} I_l^- \\ I_0^+ \end{bmatrix}. \tag{28e}$$

Explicitly, the following recurrence relation emerges:

$$Q_l = \begin{bmatrix} Q_{l,1} & Q_{l,2} \\ Q_{l,3} & Q_{l,4} \end{bmatrix} = \begin{bmatrix} Q_{l-1,1} U_{l,3} & Q_{l-1,1} U_{l,4} + Q_{l-1,2} \\ T_n U_{l,1} + R_f & T_n U_{l,2} \end{bmatrix}. \tag{29a}$$

Since scattering symmetry requires $Q_{l,1} = Q_{l,4}$ and $Q_{l,2} = Q_{l,3}$, the additional relations

$$\begin{aligned} U_{l,3} &= Q_{l,1}^{-1} T_n U_{l,2} \\ U_{l,4} &= Q_{l,1}^{-1} \left[ T_n U_{l,1} + R_f - Q_{l,2} \right] \end{aligned} \tag{29b}$$

hold and have been numerically verified, but their significance remains unclear.

Our analysis is not limited to the addition of a single node to a composite. More generally, the combined response of Eq(29a) holds for the addition of two composites of any composition-- generalized by replacing $T_n$ and $R_f$ by the responses $Q_{k,1}$ and $Q_{k,2}$ of the added $k-$ composite.

### E. Doubling and Adding
### 1. Doubling to the exiting distribution from a homogeneous node
Since the interval $h$ must be small to obtain high precision for the reflected and especially for the transmitted exiting angular Stokes vector to build a response for a homogeneous medium, it is inefficient to continually add nodes one at a time as suggested by Eqs(28). The procedure just described is the conventional diamond difference numerical solution of the transport equation with sweeps and is ineffective for high precision-- Hence, doubling.

Let the homogeneous scattering medium of width $\tau_0$ be partitioned into $2^n$ nodes $h_n$, where $n$ is a positive integer and therefore $h_n \equiv \tau_0 / 2^n$. If the first such node is denoted 1, let $R$, from Eq(22b), be the response for that node. We then find the combined response for two nodes from Eqs(28) by setting $Q_{l-1}$ to $R$. Thus, the response $Q_2$ becomes a node of width $2h_n$. Next, double the response to four nodes by combining two- two node responses again using Eqs(28) with $Q_{l-1}$ and $R$ replaced by $Q_2$. Now a node of width $4h_2$ has the response $Q_3$. This continues until the entire medium is covered to give the medium response $Q_n$. Note that the subscript of the response matrix represents the number of node doublings to that node.

Once $Q_n$ is known, the exiting intensities come from Eq(28e) with $l$ replaced by $n$

$$\begin{bmatrix} I_0^- \\ I_n^+ \end{bmatrix} = Q_n \begin{bmatrix} I_n^- \\ I_0^+ \end{bmatrix}. \tag{30}$$

The final exiting distributions at the boundaries follows by expressing the integral in the boundary condition at $\tau_0$ of Eq(1e) as

$$\int_0^1 d\mu' \mu' I(\tau_0, \mu') \simeq \sum_{m'=1}^N \omega_{m'} \mu_{m'} I_{m'}(\tau_0)$$

or in terms of $M$ and $W$ of Eqs(3e,f)

$$\int_0^1 d\mu' \mu' I(\tau_0, \mu') \simeq \underbrace{\begin{bmatrix} I_2 & I_2 & \cdots & I_2 \end{bmatrix}}_{N} MWI_n^+. \tag{31}$$

where $I_2$ is the 2 by 2 identity. Therefore, the boundary condition at the far boundary is

$$I_n^- = 2\lambda_0 D MWI_n^+, \tag{32a}$$

where $D$ is the $N$ by $N$ matrix formed by 2 by 2 blocks of the reflection matrix $D$

$$D \equiv \{D_{m,m'} \equiv D;\ m, m' = 1, \ldots, N\}. \tag{32b}$$

Since Eq(30) is

$$I_0^- = Q_{n11}I_n^- + Q_{n12}I_0^+$$
$$I_n^+ = Q_{n12}I_n^- + Q_{n11}I_0^+,$$
(33a,b)

by substituting Eq(32a) into Eqs(33b),

$$I_n^+ = [I - 2\lambda_0 Q_{n12}DMW]^{-1} Q_{n11}I_0^+$$
(34a)

and from Eq(33a)

$$I_n^- = 2\lambda_0 DMW [I - 2\lambda_0 Q_{n12}DMW]^{-1} Q_{n11}I_0^+,$$
(34b)

which when introduced into Eq(33a) gives

$$I_0^- = \{Q_{n12} + 2\lambda_0 Q_{n11}DMW[I - 2\lambda_0 Q_{n12}DMW]^{-1} Q_{n11}\}I_0^+.$$
(34c)

Equations (34a,c) define the outgoing intensity vector in terms of the incoming source vector on the near $(I_0^+)$ and far boundaries $(I_n^-)$.

We consider the interior angular intensity next.

## 2. Doubling and adding for the interior distribution

Say, one is interested in the angular Stokes 2-vector at $s+1$ interfaces of a polarizing medium as shown in Fig. 2. The Stokes vector at the $l^{th}$ interface will come from

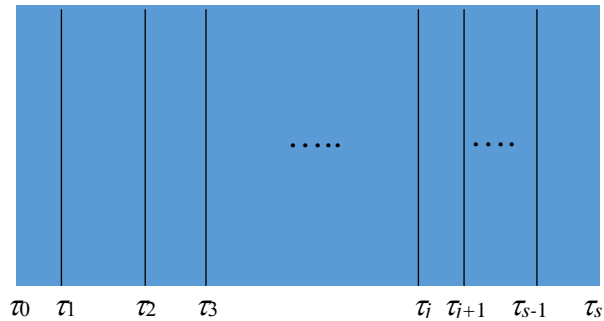

$\tau_0$ $\tau_1$ $\tau_2$ $\tau_3$ $\tau_i$ $\tau_{i+1}$ $\tau_{s-1}$ $\tau_s$
Fig. 3. Interior edits.

considering the homogeneous medium as if it were heterogeneous composed of multiple interior nodes whose surfaces correspond to desired edits. A truly heterogeneous medium of nodes of different properties is mentioned at the end of this section. Assume the surfaces are the desired interior edits uniformly distributed at intervals of $h$. Here is where the method of doubling has its advantage in that the response $R_0$ of a single node of width $h$ is known without knowledge of any intensities. As above, $R_0$ is found by doubling. Then by adding nodes using Eqs(28), the cumulative response $Q_s$ of the $s$ uniform slabs is determined. Now the subscript indicates the edit surface. Thus, the exiting distribution for the entire medium from Eq(28e) is

$$\begin{bmatrix} I_0^- \\ I_s^+ \end{bmatrix} = Q_s \begin{bmatrix} I_s^- \\ I_0^+ \end{bmatrix}. \tag{35}$$

This procedure should theoretically give the identical exiting distribution as doubling for a homogeneous medium but has the advantage of being explicit.

Finally, with boundary conditions, the exiting distributions of Eqs(34) are

$$I_s^- = 2\lambda_0 \left\{ DMW \left[ I - 2\lambda_0 Q_{s12} DMW \right]^{-1} Q_{s11} \right\} I_0^+$$
$$I_0^- = \left\{ Q_{s12} + 2\lambda_0 Q_{s11} DMW \left[ I - 2\lambda_0 Q_{s12} DMW \right]^{-1} Q_{s11} \right\} I_0^+ \tag{36a,b}$$

as given by Eqs(34b,c). In this process, $U_l$ from Eq(28b) for $l = 2,\ldots,s$ is determined and stored and used in Eq(28a) with $I_s^-$ from Eq(35) to give the angular intensity at edit $s-1$

$$\begin{bmatrix} I_{s-1}^+ \\ I_{s-1}^- \end{bmatrix} = U_s \begin{bmatrix} I_s^- \\ I_0^+ \end{bmatrix}. \tag{37a}$$

Decrementing $s$ then gives edits at surfaces $l = s-2,\ldots,2$, where $I_l^-$ comes from the previous surface to give

$$\begin{bmatrix} I_{l-1}^+ \\ I_{l-1}^- \end{bmatrix} = U_l \begin{bmatrix} I_l^- \\ I_0^+ \end{bmatrix}. \tag{37b}$$

In general, the node widths are not required to be uniform and, as demonstrated in the results section, do not all have to be of the same scattering and absorbing properties. If so, the only additional step is the response for each interior node now comes from doubling, since they could all be of different widths. Another option is to add nodes to obtain the required response for the correct node width.

**F. Reflectance and Transmission Coefficients**

Also of interest are the reflectance and transmission coefficients,

$$A^* \equiv \frac{2}{\mu_0(F_I + F_Q)} \int_0^1 d\mu' \mu' \left[ I(0,-\mu') + Q(0,-\mu') \right]$$

$$B^* \equiv \frac{2}{\mu_0(F_I + F_Q)} \int_0^1 d\mu' \mu' \left[ I(\tau_0,\mu') + Q(\tau_0,\mu') \right],$$

(38a,b)

respectively or in matrix representation

$$A^* \simeq \frac{2}{\mu_0(F_I + F_Q)} \sum_{m'=N+1}^{2N} \omega_{m'} \mu_{m'} \left[ I_{m'}(0) + Q_{m'}(0) \right]$$

$$= \frac{2}{\mu_0(F_I + F_Q)} \left[ \underbrace{\mathbf{1}_2^T \quad \mathbf{1}_2^T \quad ... \quad \mathbf{1}_2^T}_{N} \right] \mathbf{MWI}_0^-$$

(39a)

$$B^* \simeq \frac{2}{\mu_0(F_I + F_Q)} \sum_{m'=1}^{N} \omega_{m'} \mu_{m'} \left[ I_{m'}(\tau_0) + Q_{m'}(\tau_0) \right]$$

$$= \frac{2}{\mu_0(F_I + F_Q)} \left[ \underbrace{\mathbf{1}_2^T \quad \mathbf{1}_2^T \quad ... \quad \mathbf{1}_2^T}_{N} \right] \mathbf{MWI}_s^+,$$

(39b)

where $\mathbf{1}_2^T \equiv [1 \quad 1]$.

**II. Numerical Implementation**

The above theory has established an analytical representation of the fully discretized radiative transfer equation in space and direction variables for a fixed node interval and quadrature order. As observed in Eq(7), the solution requires the evaluation of a matrix exponential. In addition, we must consider the implementation of the beam source. Finally, any sound transport solution requires an internal measure of numerical convergence in spatial discretization and quadrature order. We consider each numerical implementation in turn including the special case of the grazing direction.

**A. Evaluation of the Matrix Exponential**

There are about "nineteen dubious ways" [13] to evaluate a matrix exponential. One popular representation is through matrix diagonalization, which corresponds to the ADO and response matrix-like methods [15-17] that require eigenvalues and eigenvectors of $A$. If one desires precision through simplicity, it is best to use a numerical scheme free of eigenvalues and eigenvectors. Doubling and adding is such an approach. Indeed, recent efforts in improving the ADO method concern determination of eigenvalues and eigenvectors that involve relatively sophisticated coding not always available to low-end numerical evaluators.

**1. Diamond difference (DD) approximation**

Rather than use a Taylor series approximation for the matrix exponential as above, we prefer to integrate Eq(6a) over the node

$$Y_{j+1} - Y_j = -A \int_{\tau_j}^{\tau_{j+1}} d\tau Y(\tau). \tag{40}$$

This approach opens up a large number of numerical possibilities including single and multiple step methods.

For our first method, we evaluate the integral by trapezoidal rule to give

$$Y_{j+1} = P^{*-1} P Y_j, \tag{41a}$$

with

$$P^* \equiv \left[ I + \frac{h}{2} A \right]$$

$$P \equiv \left[ I - \frac{h}{2} A \right]. \tag{41b,c}$$

Transport theorists call this the diamond difference approximation [18], but it is also a (1,1) Padé approximant for the exponential.

Equation (41a) then becomes

$$Y_{j+1} = P^{*-1} P Y_j; \qquad (42a)$$

and with partitioning

$$P^* = \begin{bmatrix} P_{11}^*(h) & P_{12}^*(h) \\ P_{21}^*(h) & P_{22}^*(h) \end{bmatrix}$$

$$P = \begin{bmatrix} P_{11}(h) & P_{12}(h) \\ P_{21}(h) & P_{22}(h) \end{bmatrix} \qquad (42b,c)$$

to give

$$\begin{bmatrix} P_{11}^* & P_{12}^* \\ P_{21}^* & P_{22}^* \end{bmatrix} \begin{bmatrix} I_{j+1}^+ \\ I_{j+1}^- \end{bmatrix} = \begin{bmatrix} P_{11} & P_{12} \\ P_{21} & P_{22} \end{bmatrix} \begin{bmatrix} I_j^+ \\ I_j^- \end{bmatrix}. \qquad (42d)$$

Thus, on rearrangement, similar to the derivation of Eq(10) to find the response matrix,

$$\begin{bmatrix} -P_{12} & P_{11}^* \\ -P_{22} & P_{21}^* \end{bmatrix} \begin{bmatrix} I_j^- \\ I_{j+1}^+ \end{bmatrix} = \begin{bmatrix} -P_{12}^* & P_{11} \\ -P_{22}^* & P_{21} \end{bmatrix} \begin{bmatrix} I_{j+1}^- \\ I_j^+ \end{bmatrix};$$

and

$$\begin{bmatrix} I_j^- \\ I_{j+1}^+ \end{bmatrix} = R \begin{bmatrix} I_{j+1}^- \\ I_j^+ \end{bmatrix},$$

where the response matrix $R$ is

$$R \equiv \begin{bmatrix} -P_{12} & P_{11}^* \\ -P_{22} & P_{21}^* \end{bmatrix}^{-1} \begin{bmatrix} -P_{12}^* & P_{11} \\ -P_{22}^* & P_{21} \end{bmatrix}. \tag{43}$$

From the definition of $A$ therefore

$$P_{11}^*(h) \equiv I + \frac{h}{2}\eta_{11}^{(1)}$$

$$P_{12}^*(h) \equiv -\frac{h}{2}\eta_{12}^{(1)}$$

$$P_{21}^*(h) \equiv \frac{h}{2}\eta_{21}^{(1)} \tag{44a,b,c,d}$$

$$P_{22}^*(h) \equiv I - \frac{h}{2}\eta_{22}^{(1)};$$

and since

$$P(h) = P^*(-h), \tag{45a}$$

we have

$$P = \begin{bmatrix} P_{22}^* & P_{21}^* \\ P_{12}^* & P_{11}^* \end{bmatrix}. \tag{45b}$$

As shown in Appendix B, it is precisely Eq(45a) that enables symmetry.

Thus, $R$ becomes

$$R \equiv \begin{bmatrix} -P_{21}^* & P_{11}^* \\ -P_{11}^* & P_{21}^* \end{bmatrix}^{-1} \begin{bmatrix} -P_{12}^* & P_{22}^* \\ -P_{22}^* & P_{12}^* \end{bmatrix}. \tag{45c}$$

The DD approximation is one of many that applies to the discrete ordinates equations. It is one of the simplest; however, as shown, not the most efficient. Most importantly, scattering symmetry is preserved.

## 2. Padé approximant

As indicated, the DD approximation is the (1,1) Padé approximant for the exponential so a natural extension is the $(q,q)$ Padé approximant

$$e^{-Ah} \simeq D_q(h)^{-1} N_q(h), \tag{46a}$$

where [13]

$$N_q(h) \equiv \left[ I + \sum_{j=1}^{q} (-1)^j \frac{(2q-j)!}{j!(2q)!} \frac{q!}{(q-j)} A^j h^j \right]$$

$$D_q(h) \equiv \left[ I + \sum_{j=1}^{q} \frac{(2q-j)!}{j!(2q)!} \frac{q!}{(q-j)} A^j h^j \right]. \tag{46b,c}$$

From Eq(46a) therefore

$$e^{-Ah} \simeq P_q^{*-1}(h) P_q(h). \tag{47}$$

Since $N_q(h) = D_q(-h)$ the $(q,q)$ Padé approximant satisfies

$$P_q^*(h) \equiv D_q(h)$$
$$P_q(h) \equiv D_q(-h), \tag{48a,b}$$

in complete agreement with the (1,1) Padé approximant, implying

$$P_q(h) \equiv P_q^*(-h) = \begin{bmatrix} P_{q,22}^* & P_{q,21}^* \\ P_{q,12}^* & P_{q,11}^* \end{bmatrix}.$$

to give the identical form of response matrix

$$R \equiv \begin{bmatrix} -P_{q,21}^* & P_{q,11}^* \\ -P_{q,11}^* & P_{q,21}^* \end{bmatrix}^{-1} \begin{bmatrix} -P_{q,12}^* & P_{q,22}^* \\ -P_{q,22}^* & P_{q,12}^* \end{bmatrix}. \tag{49}$$

found in the DD approximation. Thus, we have established a general Padé approximant option. However, the (1,1) approximant seems to be the most efficient.

### 3. RK5 approximation

Since one finds the response matrix by numerically solving a first order ODE, numerous numerical methods become possible. For example, we now try the RK5 [19] solver of Fehlberg since it has proven to be generally quite successful and forms the basis of MATLAB$^{TM}$'s ODE45 solver.

Before applying vector RK5, the exponential is recast in the appropriate form to maintain symmetry of the response matrix, *i.e.*,

$$e^{-Ah} = \left[e^{Ah/2}\right]^{-1}\left[e^{-Ah/2}\right] = \boldsymbol{P}^{*-1}\boldsymbol{P} \tag{50a}$$

with

$$\begin{aligned}\boldsymbol{P}^*(h) &\equiv e^{Ah/2} \\ \boldsymbol{P}(h) &\equiv e^{-Ah/2};\end{aligned} \tag{50b}$$

and obviously $\boldsymbol{P}(h) = \boldsymbol{P}^*(-h)$ maintains symmetry.

Next, RK5 is applied to the ODE

$$\frac{d}{d\tau}\boldsymbol{\xi}(\tau) - (\boldsymbol{A}/2)\boldsymbol{\xi}(\tau) = 0 \tag{51a}$$

with solution

$$\boldsymbol{\xi}_{j+1} = e^{Ah/2}\boldsymbol{\xi}_j \tag{51b}$$

to estimate $e^{Ah/2}$ by RK5, which is [19]

$$\xi_{j+1} = \xi_j + \sum_{i=1}^{6} b_i k_i$$

$$k_i = \frac{h}{2} A \left( \xi_j + \sum_{n=1}^{i-1} a_{in} k_n \right), \; i = 1,...,6.$$

(51c,d)

Table 1 gives all coefficients in Eq(51c,d).

### Table 1. Butcher Table for RK5.

| | | | | | | | | | | | |
|---|---|---|---|---|---|---|---|---|---|---|---|
| 1/4 | | | | | | $a_{12}$ | | | | | |
| 3/32 | 9/32 | | | | | $a_{13}$ | $a_{23}$ | | | | |
| 1932/2197 | −7200/2197 | 7296/2197 | | | | $a_{14}$ | $a_{24}$ | $a_{34}$ | | | |
| 439/216 | −8 | 3680/513 | −845/4104 | | | $a_{15}$ | $a_{25}$ | $a_{35}$ | $a_{45}$ | | |
| −8/27 | 2 | −3544/2565 | 1859/4104 | −11/40 | | $a_{16}$ | $a_{26}$ | $a_{36}$ | $a_{46}$ | $a_{56}$ | |
| 16/135 | 0 | 6656/12825 | 28561/56430 | −9/50 | 2/55 | $b_1$ | 0 | $b_3$ | $b_4$ | $b_5$ | $b_6$ |

If

$$k_i = g_i(h) \xi_j,$$

then

$$e^{Ah/2} \simeq P^*(h) = P(-h) \equiv I + \sum_{i=1}^{6} b_i g_i(h),$$

(52a)

where

$$g_i = \frac{h}{2} A \left[ I + \sum_{i=1}^{i-1} a_{in} g_n(h) \right].$$

(52b)

Since

$$P^*(h) = \begin{bmatrix} P_{11}^* & P_{12}^* \\ P_{21}^* & P_{22}^* \end{bmatrix};$$

(53a)

on further reduction

$$P^*(h) = I + \sum_{k=1}^{6} e_i h^k A^k,  \qquad (53b)$$

which is a Taylor polynomial in $hA$ giving the same decomposition in even and odd powers as the Taylor series of Eq(19) implying

$$P = \begin{bmatrix} P^*_{22} & P^*_{21} \\ P^*_{12} & P^*_{11} \end{bmatrix},$$

and subsequently from Appendix B

$$R \equiv \begin{bmatrix} -P^*_{21} & P^*_{11} \\ -P^*_{11} & P^*_{21} \end{bmatrix}^{-1} \begin{bmatrix} -P^*_{12} & P^*_{22} \\ -P^*_{22} & P^*_{12} \end{bmatrix}. \qquad (54)$$

Not surprisingly, the RK5 option proves to be the most efficient.

### B. Implementing a Beam Source

The inclusion of a beam source in transport calculations is most commonly accommodated through a particular solution. This requires the intensity to be represented by an uncollided and a collided component having had at least one collision. The uncollided component is determined analytically and becomes a volume source in the radiative transfer equation for the collided component. A particular solution is then sought for the volume source. In general however, an analytical particular solution can be rather elusive and if found, for example as an integration over the Green's function, is as extensive a calculation as the original. More importantly, a fixed source in the transport equation destroys the advantage of doubling since now two nodes may not be alike. In this case, the response for a homogeneous medium cannot be found independently of the intensity. Thus, we seek an alternative.

Since the source is emitted in a direction $\mu_0$ and when integrated over all directions is normalized to

$$\frac{1}{2}\begin{bmatrix} F_I & F_Q \end{bmatrix}^T,$$

we introduce a large number, say $1/\varepsilon$, as the source strength and $\varepsilon$ at the quadrature weight and include $\mu_0$ in the quadrature list. Thus, while the source is large it remains normalized and only enters as an integrated contribution. From experimentation, $\varepsilon$ should be chosen larger $10^{-10}$; but to be safe, we choose $10^{-100}$ here. Both give indistinguishable results. This procedure has the added advantage of enabling beam directions other than quadrature directions.

## C. Special Case of the Grazing Direction ($\mu$=0)

The grazing direction, $\mu = 0$, is always problematic in transport calculations since it leads to a singular solution in which the derivative with respect to $\mu$ at the free surfaces is undefined. This case is most easily treated by setting the first term in Eq(1a) to zero when $\mu = 0^{\pm}$ giving

$$I(\tau, 0^{\pm}) = \frac{1}{2}\omega \int_{-1}^{1} d\mu' Q(0) Q^T(\mu') I(\tau, \mu'). \tag{55}$$

In vector notation

$$I(\tau_j, 0^{\pm}) = \frac{1}{2}\omega Q(0) Q^T W \left[ I_j^+ + I_j^- \right] \tag{56a}$$

with

$$Q^T \equiv diag\left\{ Q^T(\mu_m); \; m = 1,...,N \right\}. \tag{56b}$$

Except when $j = 0, s$, the intensity is continuous at the grazing angle.

## D. Convergence

We now discuss the concept of convergence of the solution with regard to spatial discretization and angular quadrature. A notable feature of this investigation is that we include convergence as an integral part of the solution.

The solution found thus far is for a fixed node interval $h$ and quadrature order $N$. For consistency, the true solution is found when $h$ approaches zero and the quadrature order approaches infinity. Therefore, the true (infinite dimensional) angular Stokes 2-vector at an interface $l$ is

$$\begin{bmatrix} \boldsymbol{I}_l^{+T} & \boldsymbol{I}_l^{-T} \end{bmatrix}^T = \lim_{\substack{h \to 0 \\ N \to \infty}} \begin{bmatrix} \boldsymbol{I}_l^{+T}(h, N) & \boldsymbol{I}_l^{-T}(h, N) \end{bmatrix}^T. \tag{57}$$

A sensitivity analysis, performed along with doubling and adding procedures enables the approximation of the limit. That is, for each quadrature order $N$, a sequence of solutions for node intervals $h/2^l$ from $l = l_{ini}$ to convergence (to be described) becomes an integral part of the solution. On convergence, the quadrature order is incremented from $N_{ini}$ by $N_{inc}$ until the solution sequences for both discretizations converge to a desired precision.

Convergence, as used here, is "engineering convergence" or the relative error of the reflectance and transmittance or angular Stokes vector components between two sequence elements defined by $l$ and $N$. To achieve convergence, one applies a "faux" quadrature, which amounts to including $L_e$ edit directions in the quadrature direction list all with zero weight. In this way, the transport equation itself naturally interpolates the solution at the edits without influencing the discrete balance. Thus, there is no need for an outside interpolation. Moreover, the solution at the edits will have the same error as dictated by the spatial and quadrature approximations  The edits appear in each quadrature approximation $N$ and therefore provide a common basis for comparison of overall error at the $s$ edit interfaces. Convergence occurs when the maximum relative error of consecutive spatially converged quadrature approximations for all edits fall below a desired relative error. A similar procedure has been applied in [10] as a post-processing step once the solution at the quadrature abscissae has been established. The disadvantage of faux edits is that the size of the response matrix is increased by the number of edits since now one must consider $2(N+L_e+1)$ quadrature abscissae rather than $2(N+1)$, where the additional 1 is for the grazing direction. Generally, the disadvantage of a larger response matrix is more than offset by the simplicity of supplying edits through input with virtually no additional programming-- unless an enormous number if edits are required.

The convergence procedure just described, called original (O) convergence, is essentially the convergence of the (O) sequence

$$\left[ \mathbf{I}_l^{+T}(h_l, N) \quad \mathbf{I}_l^{-T}(h_l, N) \right]^T \tag{58}$$

as $(h_l, N) \to (0, \infty)$ respectively. Simultaneously, we apply convergence acceleration (CA) in the form of Wynn-epsilon (*W-e*) and Richardson (*R*) acceleration (or extrapolation) [20] to the O sequence. In convergence acceleration, one attempts to capture the asymptotic behavior in both *l* and *N*. Hence, convergence of the original, *W-e* and *R* sequences are monitored and whichever converges first is the converged solution. While sometimes unsuccessful, the CA algorithms will converge as quickly as or slightly less quickly than the original, but nevertheless will eventually converge to the same limit to provide additional assurance.

### III. Numerical Comparisons and Benchmarks

We finally come to numerical results. Two examples of scattering matrix factors $\mathbf{Q}(\mu)$ and reflecting boundaries will demonstrate the precision of the theory of doubling and adding and confirm its numerical implementation. The examples, from the literature, provide direct assessment of the precision of the method as presented.

For the first comparison (C1), the scattering matrix factor is [9]

$$\mathbf{Q}(\mu) \equiv \frac{3}{2}(c+2)^{-1/2} \begin{bmatrix} c\mu^2 + \frac{2}{3}(1-c) & (2c)^{1/2}(1-\mu^2) \\ \frac{1}{2}(c+2) & 0 \end{bmatrix}, \tag{59a}$$

where *c* represents the split between Rayleigh (*c* =1) and isotropic (*c* = 0) scattering. The reflection matrix is

$$\mathbf{D} \equiv \begin{bmatrix} 1 & 1 \\ 1 & 1 \end{bmatrix}. \tag{59b}$$

For the second comparison (C2),

$$Q(\mu) \equiv \begin{bmatrix} 1 & \left(\dfrac{c}{8}\right)^{1/2}(1-3\mu^2) \\ 0 & 3\left(\dfrac{c}{8}\right)^{1/2}(1-\mu^2) \end{bmatrix}, \quad (60a)$$

and

$$D \equiv \begin{bmatrix} 1 & 0 \\ 0 & 0 \end{bmatrix}. \quad (60b)$$

For all reported results, the sequence in $l$ begins with $l_{ini} = 5$ in order to establish a small enough node interval for a reasonable initial approximation and advances by unity. The sequence in $N$ begins at $N_{ini} = 16$ and advances by $N_{inc} = 4$. All results designated as benchmarks quote seven or more places (eight or more significant figures). Finally all cases are for a perpendicular beam ($\mu_0 = 1$)

### A. Comparison C1
We now consider the five cases of Ref. 9 given in Table 2. The source is

**Table 2a. Cases for C1.**

| Case | $\omega$ | $c$ | $\lambda_0$ | $\tau_0$ | $\mu_0$ |
|------|------|-----|-------|------|-------|
| 1 | 0.9 | 1.0 | 0.0 | 1.0 | 1.0 |
| 2 | 0.9 | 1.0 | 0.05 | 1.0 | 1.0 |
| 3 | 0.9 | 0.8 | 0.05 | 5.0 | 1.0 |
| 4 | 0.9 | 0.8 | 0.1 | 5.0 | 1.0 |
| 5 | 0.9 | 0.8 | 0.0 | $\infty$ | 1.0 |

$F_I = F_Q = 1/2$ and the node response matrix is from the DD algorithm [Eqs(44), Eq(45b)]. The thickness for the last case is 100 optical depths to simulate a semi-infinite medium. Table 2b displays the reflectance and transmittance given by Eqs(39a,b) to 9 places for the original (*O*) sequence and the Wynn-epsilon (*W-e*) and Richardsons (*R*) accelerations.

**Table 2b. Reflectance and transmittance for C1.**

| | $R_f$ | | | $T_n$ | | |
|---|---|---|---|---|---|---|
| Case | O | W-e | R | O | W-e | R |
| 1 | 0.270229314 | 0.270229314 | 0.270229314 | 0.596165717 | 0.596165717 | 0.596165717 |
| 2 | 0.299571235 | 0.299571235 | 0.299571235 | 0.617990132 | 0.617990132 | 0.617990132 |
| 3 | 0.417536585 | 0.417536585 | 0.417536585 | 0.082957853 | 0.082957853 | 0.082957853 |
| 4 | 0.418031312 | 0.418031312 | 0.418031312 | 0.087333753 | 0.087333753 | 0.087333753 |
| 5 | 0.419612544 | 0.419612544 | 0.419612544 | 2.57396e-23 | 2.57396e-23 | 2.57396e-23 |

As one observes, all solution sequences for each case converge to the identical value, with convergence specifics given in Table 2c. The mode of convergence in quadrature for all cases is acceleration [*W-e*(2) or *R*(3)] confirming the value of acceleration. Note that the transmittance for the last case is a number representing zero. On rounding, all values agree to the five places quoted in Ref. 9.

With regard to convergence in node discretization, Table 3 shows statistics for convergence in $l$ at convergence of the quadrature order ($N = 40$) for Case 1. The relative error (columns 6-8), based on the angular Stokes vector for four values of $\mu$

**Table 2c. Case statistics for DD response Matrix.**

| Case | N | Mode | l | h | Time(s) |
|---|---|---|---|---|---|
| 1 | 40 | 3 | 10 | 9.7656E-02 | 2.5469E+00 |
| 2 | 40 | 2 | 10 | 9.7656E-02 | 2.5625E+00 |
| 3 | 32 | 2 | 12 | 2.4414E-02 | 1.9688E+00 |
| 4 | 32 | 2 | 12 | 2.4414E-02 | 1.9219E+00 |
| 5 | 32 | 2 | 17 | 7.6294E-04 | 4.2188E+00 |

**Table 2d. Case statistics for RK5 response Matrix.**

| Case | N | Mode | l | h | Time(s) |
|---|---|---|---|---|---|
| 1 | 40 | 2 | 8 | 3.9062E-01 | 1.4531E+00 |
| 2 | 40 | 2 | 8 | 3.9062E-01 | 1.4219E+00 |
| 3 | 32 | 2 | 8 | 3.9062E-01 | 6.7188E-01 |
| 4 | 32 | 2 | 8 | 3.9062E-01 | 7.1875E-01 |

($\pm 1, 0^{\pm}$), reflectance and transmittance, confirms the advantage of acceleration by five orders of magnitude. Columns 3 to 5 indicate that at convergence in $l$ ($l = 10$), the intensity of the original sequence has yet to converged.

**Table 3. Convergence in $l$ for C1.**

| N | l | O | W-e | R | Rel:O | Rel:W-e | Rel:R |
|---|---|---|---|---|---|---|---|
| 84 | 7 | 0 | 0 | 0 | 1.09E+00 | 1.09E+00 | 1.09E+00 |
| 84 | 8 | 0 | 0 | 0 | 2.54E-06 | 2.54E-06 | 2.54E-06 |
| 84 | 9 | 0 | 0 | 0 | 6.36E-07 | 8.48E-07 | 8.48E-07 |
| 84 | 10 | 0 | 8 | 8 | 1.59E-07 | 1.29E-12 | 5.14E-12 |

Tables 2 and 3 have been reproduced for the response matrix as determined by RK5. Included in Tables 2a,b are the times of computation showing nearly a factor of two advantage of RK5. The advantage obviously comes from more rapid convergence in $l$.

The final comparison of this section is for the angular intensity. Tables 4a-c provide the angular intensities for cases 1, 3 and 5 corresponding to the 4-place results of Ref. 9. The RK5 algorithm gives the response matrix, which will be the preferred algorithm. In the tables, numbers on the order of $10^{-16}$, $10^{-17}$ and $10^{-24}$ approximate zero. Uncharacteristic of a Siewert benchmark, on rounding several entries (shaded) seem not to agree. The discrepancies occur near the beam direction and the grazing angle. The 7 places quoted in the tables are believed to be correct to better than one unit in the last place and therefore represent a highly precise benchmark.

**Table 4a. Angular Intensity Surrogates for Case 1.**

| $\mu$ | $I(0,\mu)+Q(0,\mu)$ | $I(0,\mu)-Q(0,\mu)$ | $I(\tau_0,\mu)+Q(\tau_0,\mu)$ | $I(\tau_0,\mu)-Q(\tau_0,\mu)$ |
|---|---|---|---|---|
| -1.0000000E+00 | 2.7270983E-01 | 1.38777E-16 | 0.0000000E+00 | 0.0000000E+00 |
| -9.8000000E-01 | 2.7175139E-01 | -4.4270215E-03 | 0.0000000E+00 | 0.0000000E+00 |
| -9.6000000E-01 | 2.7084890E-01 | -8.8792557E-03 | 0.0000000E+00 | 0.0000000E+00 |
| -9.2000000E-01 | 2.6921872E-01 | -1.7861875E-02 | 0.0000000E+00 | 0.0000000E+00 |
| -9.0000000E-01 | 2.6849465E-01 | -2.2393513E-02 | 0.0000000E+00 | 0.0000000E+00 |
| -8.4000000E-01 | 2.6670890E-01 | -3.6157081E-02 | 0.0000000E+00 | 0.0000000E+00 |
| -7.2000000E-01 | 2.6505129E-01 | -6.4491514E-02 | 0.0000000E+00 | 0.0000000E+00 |
| -6.4000000E-01 | 2.6552846E-01 | -8.4007853E-02 | 0.0000000E+00 | 0.0000000E+00 |
| -5.2000000E-01 | 2.6883786E-01 | -1.1416969E-01 | 0.0000000E+00 | 0.0000000E+00 |
| -4.0000000E-01 | 2.7511512E-01 | -1.4496774E-01 | 0.0000000E+00 | 0.0000000E+00 |
| -3.2000000E-01 | 2.8032793E-01 | -1.6519930E-01 | 0.0000000E+00 | 0.0000000E+00 |
| -2.8000000E-01 | 2.8289390E-01 | -1.7492871E-01 | 0.0000000E+00 | 0.0000000E+00 |
| -2.0000000E-01 | 2.8691785E-01 | -1.9282241E-01 | 0.0000000E+00 | 0.0000000E+00 |
| -1.6000000E-01 | 2.8788517E-01 | -2.0062081E-01 | 0.0000000E+00 | 0.0000000E+00 |
| -1.0000000E-01 | 2.8746655E-01 | -2.1045043E-01 | 0.0000000E+00 | 0.0000000E+00 |
| -6.0000000E-02 | 2.8558155E-01 | -2.1547366E-01 | 0.0000000E+00 | 0.0000000E+00 |
| -2.0000000E-02 | 2.8120468E-01 | -2.1842411E-01 | 0.0000000E+00 | 0.0000000E+00 |
| 2.0000000E-02 | 0.0000000E+00 | 0.0000000E+00 | 1.5519902E-01 | -1.0065083E-01 |
| 6.0000000E-02 | 0.0000000E+00 | 0.0000000E+00 | 1.6699562E-01 | -1.0617196E-01 |
| 1.0000000E-01 | 0.0000000E+00 | 0.0000000E+00 | 1.7760632E-01 | -1.1091991E-01 |
| 1.6000000E-01 | 0.0000000E+00 | 0.0000000E+00 | 1.9186393E-01 | -1.1653577E-01 |
| 2.0000000E-01 | 0.0000000E+00 | 0.0000000E+00 | 1.9976602E-01 | -1.1869661E-01 |
| 2.8000000E-01 | 0.0000000E+00 | 0.0000000E+00 | 2.1107950E-01 | -1.1830289E-01 |
| 3.2000000E-01 | 0.0000000E+00 | 0.0000000E+00 | 2.1481856E-01 | -1.1592202E-01 |
| 4.0000000E-01 | 0.0000000E+00 | 0.0000000E+00 | 2.1977352E-01 | -1.0782175E-01 |
| 5.2000000E-01 | 0.0000000E+00 | 0.0000000E+00 | 2.2408816E-01 | -9.0203735E-02 |
| 6.4000000E-01 | 0.0000000E+00 | 0.0000000E+00 | 2.2769890E-01 | -6.9135568E-02 |
| 7.2000000E-01 | 0.0000000E+00 | 0.0000000E+00 | 2.3052064E-01 | -5.4166547E-02 |
| 8.4000000E-01 | 0.0000000E+00 | 0.0000000E+00 | 2.3582472E-01 | -3.1100052E-02 |
| 9.2000000E-01 | 0.0000000E+00 | 0.0000000E+00 | 2.4014893E-01 | -1.5558841E-02 |
| 9.6000000E-01 | 0.0000000E+00 | 0.0000000E+00 | 2.4254720E-01 | -7.7775510E-03 |
| 9.8000000E-01 | 0.0000000E+00 | 0.0000000E+00 | 2.4380406E-01 | -3.8879041E-03 |
| 1.0000000E+00 | 0.0000000E+00 | 0.0000000E+00 | 2.4509872E-01 | 4.16334E-17 |

### Table 4b. Angular Intensity Surrogates for Case 3.

| $\mu$ | $I(0,\mu)+Q(0,\mu)$ | $I(0,\mu)-Q(0,\mu)$ | $I(\tau_0,\mu)+Q(\tau_0,\mu)$ | $I(\tau_0,\mu)-Q(\tau_0,\mu)$ |
|---|---|---|---|---|
| -1.0000000E+00 | 4.3766545E-01 | 0.0000000E+00 | 8.2957860E-03 | 0.0000000E+00 |
| -9.8000000E-01 | 4.3623750E-01 | -3.8559344E-03 | 8.2957860E-03 | 0.0000000E+00 |
| -9.6000000E-01 | 4.3483442E-01 | -7.7025397E-03 | 8.2957860E-03 | 0.0000000E+00 |
| -9.2000000E-01 | 4.3210255E-01 | -1.5366684E-02 | 8.2957860E-03 | 0.0000000E+00 |
| -9.0000000E-01 | 4.3077352E-01 | -1.9183656E-02 | 8.2957860E-03 | 0.0000000E+00 |
| -8.4000000E-01 | 4.2693176E-01 | -3.0570540E-02 | 8.2957860E-03 | 0.0000000E+00 |
| -7.2000000E-01 | 4.1987237E-01 | -5.3025167E-02 | 8.2957860E-03 | 0.0000000E+00 |
| -6.4000000E-01 | 4.1557918E-01 | -6.7723457E-02 | 8.2957860E-03 | 0.0000000E+00 |
| -5.2000000E-01 | 4.0958836E-01 | -8.9280747E-02 | 8.2957860E-03 | 0.0000000E+00 |
| -4.0000000E-01 | 4.0373233E-01 | -1.1010934E-01 | 8.2957860E-03 | 0.0000000E+00 |
| -3.2000000E-01 | 3.9948651E-01 | -1.2347521E-01 | 8.2957860E-03 | 0.0000000E+00 |
| -2.8000000E-01 | 3.9709647E-01 | -1.2996304E-01 | 8.2957860E-03 | 0.0000000E+00 |
| -2.0000000E-01 | 3.9129686E-01 | -1.4244408E-01 | 8.2957860E-03 | 0.0000000E+00 |
| -1.6000000E-01 | 3.8756620E-01 | -1.4837189E-01 | 8.2957860E-03 | 0.0000000E+00 |
| -1.0000000E-01 | 3.8012159E-01 | -1.5671432E-01 | 8.2957860E-03 | 0.0000000E+00 |
| -6.0000000E-02 | 3.7303389E-01 | -1.6174208E-01 | 8.2957860E-03 | 0.0000000E+00 |
| -2.0000000E-02 | 3.6229050E-01 | -1.6599621E-01 | 8.2957860E-03 | 0.0000000E+00 |
| 2.0000000E-02 | 0.0000000E+00 | 0.0000000E+00 | 3.4429517E-02 | -4.4339364E-03 |
| 6.0000000E-02 | 0.0000000E+00 | 0.0000000E+00 | 3.6862585E-02 | -4.4568618E-03 |
| 1.0000000E-01 | 0.0000000E+00 | 0.0000000E+00 | 3.9118461E-02 | -4.4985251E-03 |
| 1.6000000E-01 | 0.0000000E+00 | 0.0000000E+00 | 4.2415844E-02 | -4.5681549E-03 |
| 2.0000000E-01 | 0.0000000E+00 | 0.0000000E+00 | 4.4615859E-02 | -4.6125199E-03 |
| 2.8000000E-01 | 0.0000000E+00 | 0.0000000E+00 | 4.9114221E-02 | -4.6848868E-03 |
| 3.2000000E-01 | 0.0000000E+00 | 0.0000000E+00 | 5.1440191E-02 | -4.7091670E-03 |
| 4.0000000E-01 | 0.0000000E+00 | 0.0000000E+00 | 5.6297073E-02 | -4.7250663E-03 |
| 5.2000000E-01 | 0.0000000E+00 | 0.0000000E+00 | 6.4193837E-02 | -4.6296908E-03 |
| 6.4000000E-01 | 0.0000000E+00 | 0.0000000E+00 | 7.2876210E-02 | -4.2892805E-03 |
| 7.2000000E-01 | 0.0000000E+00 | 0.0000000E+00 | 7.9071994E-02 | -3.8431081E-03 |
| 8.4000000E-01 | 0.0000000E+00 | 0.0000000E+00 | 8.8882540E-02 | -2.6987565E-03 |
| 9.2000000E-01 | 0.0000000E+00 | 0.0000000E+00 | 9.5708841E-02 | -1.5370795E-03 |
| 9.6000000E-01 | 0.0000000E+00 | 0.0000000E+00 | 9.9194538E-02 | -8.1806117E-04 |
| 9.8000000E-01 | 0.0000000E+00 | 0.0000000E+00 | 1.0095393E-01 | -4.2170186E-04 |
| 1.0000000E+00 | 0.0000000E+00 | 0.0000000E+00 | 1.0272381E-01 | 0.0000000E+00 |

### Table 4c. Angular Intensity Surrogates for Case 5.

| $\mu$ | $I(0,\mu)+Q(0,\mu)$ | $I(0,\mu)-Q(0,\mu)$ | $I(\tau_0,\mu)+Q(\tau_0,\mu)$ | $I(\tau_0,\mu)-Q(\tau_0,\mu)$ |
|---|---|---|---|---|
| -1.0000000E+00 | 4.4069289E-01 | 0.0000000E+00 | 5.1479360E-24 | 0.0000000E+00 |
| -9.8000000E-01 | 4.3919308E-01 | -3.8624118E-03 | 5.1479360E-24 | 0.0000000E+00 |
| -9.6000000E-01 | 4.3771928E-01 | -7.7151307E-03 | 5.1479360E-24 | 0.0000000E+00 |
| -9.2000000E-01 | 4.3484948E-01 | -1.5390472E-02 | 5.1479360E-24 | 0.0000000E+00 |
| -8.4000000E-01 | 4.2941781E-01 | -3.0613015E-02 | 5.1479360E-24 | 0.0000000E+00 |
| -7.2000000E-01 | 4.2200715E-01 | -5.3088077E-02 | 5.1479360E-24 | 0.0000000E+00 |
| -6.4000000E-01 | 4.1750706E-01 | -6.7796055E-02 | 5.1479360E-24 | 0.0000000E+00 |
| -5.2000000E-01 | 4.1124384E-01 | -8.9363449E-02 | 5.1479360E-24 | 0.0000000E+00 |
| -4.0000000E-01 | 4.0515266E-01 | -1.1019806E-01 | 5.1479360E-24 | 0.0000000E+00 |
| -3.2000000E-01 | 4.0076510E-01 | -1.2356624E-01 | 5.1479360E-24 | 0.0000000E+00 |
| -2.8000000E-01 | 3.9830744E-01 | -1.3005486E-01 | 5.1479360E-24 | 0.0000000E+00 |
| -2.0000000E-01 | 3.9237708E-01 | -1.4253698E-01 | 5.1479360E-24 | 0.0000000E+00 |
| -1.6000000E-01 | 3.8858236E-01 | -1.4846521E-01 | 5.1479360E-24 | 0.0000000E+00 |
| -1.0000000E-01 | 3.8104144E-01 | -1.5680841E-01 | 5.1479360E-24 | 0.0000000E+00 |
| -6.0000000E-02 | 3.7388754E-01 | -1.6183705E-01 | 5.1479360E-24 | 0.0000000E+00 |
| -2.0000000E-02 | 3.6307236E-01 | -1.6609293E-01 | 5.1479360E-24 | 0.0000000E+00 |

Table 5 provides the convergence specifics for Tables 4a-c. All values converged by *W-e* in *l* and *R* in *N*.

**Table 5. Convergence specifics for Tables 4a-c.**

| N | Mode | l | h | time(s) |
|---|---|---|---|---|
| 52 | 3 | 8 | 3.9062E-03 | 7.3125E+00 |
| 52 | 3 | 8 | 1.9531E-02 | 7.6719E+00 |
| 52 | 3 | 12 | 2.4414E-02 | 2.6828E+01 |

Finally, the angular intensities of Tables 4a-c have been reproduced by the response matrix determined by a (21,21) Padé approximant yielding the identical tables, but requiring a factor of three more computational time.

## B. Comparison C2

Table 6 gives the two homogeneous media cases we first consider as found in Ref. 10. The source normalizations are $F_I = 1$, $F_Q = 0.8$.

**Table 6. Cases for C2.**

| Case | $\omega$ | $c$ | $\lambda_0$ | $\tau_0$ | $\mu_0$ |
|---|---|---|---|---|---|
| 1 | 0.99 | 0.5 | 0.2 | 2.0 | 1.0 |
| 2 | 1.0 | 0.5 | 1.0 | 1.0 | 1.0 |

Tables 7a,b give the two components of the intensity vector for case 1 for three interior positions and the free surfaces. On rounding all values are in agreement with the 4 places of Ref. 10. All entries in the tables are expected to be precise to better than one unit in the last place. This benchmark required less than 6*s* on a LENOVO YOGA (2.4GHz) PC and converged for (*N*,*l*) = (40,6).

Tables 8a,b provide a new benchmark. In particular, this benchmark is for the medium of Case 1 without a trace of absorption, *i.e.*, the conservative case. It should be noted that this is a true conservative case where $\omega$ is unity, unlike similar methods that can treat only $\omega$ near unity. Here, no special adaptation of the method is necessary— an advantage of doubling and adding. Like the previous benchmark, all entries are expected to be precise to better than one unit in the last place and the convergence and computational effort are the same as for tables 7.

## Table 7a. Component $I(\eta\tau,\mu)$ Case 1.

| $\mu\backslash\eta$ | 0.0 | 0.1 | 0.5 | 0.75 | 1.0 |
|---|---|---|---|---|---|
| -1.0E+00 | 5.1625126E-01 | 4.9253619E-01 | 3.2822316E-01 | 2.1140219E-01 | 1.0578024E-01 |
| -9.0E-01 | 5.2315497E-01 | 5.0263004E-01 | 3.4015598E-01 | 2.1891330E-01 | 1.0578024E-01 |
| -8.0E-01 | 5.3051875E-01 | 5.1370485E-01 | 3.5397895E-01 | 2.2798243E-01 | 1.0578024E-01 |
| -7.0E-01 | 5.3801916E-01 | 5.2559158E-01 | 3.6999757E-01 | 2.3906084E-01 | 1.0578024E-01 |
| -6.0E-01 | 5.4511721E-01 | 5.3794390E-01 | 3.8851866E-01 | 2.5277528E-01 | 1.0578024E-01 |
| -5.0E-01 | 5.5096233E-01 | 5.5014324E-01 | 4.0976471E-01 | 2.7000095E-01 | 1.0578024E-01 |
| -4.0E-01 | 5.5428988E-01 | 5.6119768E-01 | 4.3366185E-01 | 2.9193107E-01 | 1.0578024E-01 |
| -3.0E-01 | 5.5335137E-01 | 5.6972078E-01 | 4.5938060E-01 | 3.1998933E-01 | 1.0578024E-01 |
| -2.0E-01 | 5.4580371E-01 | 5.7413534E-01 | 4.8467317E-01 | 3.5478555E-01 | 1.0578024E-01 |
| -1.0E-01 | 5.2752246E-01 | 5.7274026E-01 | 5.0664341E-01 | 3.9109987E-01 | 1.0578024E-01 |
| 0.0E+00 | 4.7883405E-01 | 5.6188201E-01 | 5.2541584E-01 | 4.1833603E-01 | 1.0578024E-01 |
| 0.0E+00 | 0.0000000E+00 | 5.6188201E-01 | 5.2541584E-01 | 4.1833603E-01 | 2.6389096E-01 |
| 1.0E-01 | 0.0000000E+00 | 4.7037849E-01 | 5.4088071E-01 | 4.4218949E-01 | 3.0218930E-01 |
| 2.0E-01 | 0.0000000E+00 | 3.4231196E-01 | 5.4775350E-01 | 4.6301449E-01 | 3.3215483E-01 |
| 3.0E-01 | 0.0000000E+00 | 2.6478651E-01 | 5.3846405E-01 | 4.7726335E-01 | 3.5792024E-01 |
| 4.0E-01 | 0.0000000E+00 | 2.1625395E-01 | 5.1794682E-01 | 4.8256301E-01 | 3.7816525E-01 |
| 5.0E-01 | 0.0000000E+00 | 1.8372393E-01 | 4.9321159E-01 | 4.8041977E-01 | 3.9225004E-01 |
| 6.0E-01 | 0.0000000E+00 | 1.6072624E-01 | 4.6829681E-01 | 4.7348858E-01 | 4.0087560E-01 |
| 7.0E-01 | 0.0000000E+00 | 1.4381503E-01 | 4.4502921E-01 | 4.6397648E-01 | 4.0529104E-01 |
| 8.0E-01 | 0.0000000E+00 | 1.3101216E-01 | 4.2408279E-01 | 4.5338139E-01 | 4.0672523E-01 |
| 9.0E-01 | 0.0000000E+00 | 1.2110766E-01 | 4.0558986E-01 | 4.4263435E-01 | 4.0618176E-01 |
| 1.0E+00 | 0.0000000E+00 | 1.1332208E-01 | 3.8944521E-01 | 4.3228236E-01 | 4.0441481E-01 |

## Table 7b. Component $Q(\eta\tau,\mu)$ Case 1.

| $\mu\backslash\eta$ | 0.0 | 0.1 | 0.5 | 0.75 | 1.0 |
|---|---|---|---|---|---|
| -1.0E+00 | 0.0000000E+00 | 0.0000000E+00 | 0.0000000E+00 | 0.0000000E+00 | 0.0000000E+00 |
| -9.0E-01 | -9.8423476E-03 | -7.9047423E-03 | -3.2507254E-03 | -1.5847710E-03 | 0.0000000E+00 |
| -8.0E-01 | -1.9762869E-02 | -1.5884048E-02 | -6.6045032E-03 | -3.2785508E-03 | 0.0000000E+00 |
| -7.0E-01 | -2.9761000E-02 | -2.3935245E-02 | -1.0073185E-02 | -5.1108832E-03 | 0.0000000E+00 |
| -6.0E-01 | -3.9833764E-02 | -3.2050540E-02 | -1.3664795E-02 | -7.1213136E-03 | 0.0000000E+00 |
| -5.0E-01 | -4.9977525E-02 | -4.0216216E-02 | -1.7376088E-02 | -9.3617639E-03 | 0.0000000E+00 |
| -4.0E-01 | -6.0195292E-02 | -4.8415964E-02 | -2.1177023E-02 | -1.1894717E-02 | 0.0000000E+00 |
| -3.0E-01 | -7.0516089E-02 | -5.6646163E-02 | -2.4983876E-02 | -1.4769531E-02 | 0.0000000E+00 |
| -2.0E-01 | -8.1029782E-02 | -6.4950009E-02 | -2.8644416E-02 | -1.7902694E-02 | 0.0000000E+00 |
| -1.0E-01 | -9.1931846E-02 | -7.3436948E-02 | -3.2087780E-02 | -2.0663050E-02 | 0.0000000E+00 |
| 0.0E+00 | -1.0413252E-01 | -8.2268327E-02 | -3.5551618E-02 | -2.2484415E-02 | 0.0000000E+00 |
| 0.0E+00 | 0.0000000E+00 | -8.2268327E-02 | -3.5551618E-02 | -2.2484415E-02 | -1.8373501E-02 |
| 1.0E-01 | 0.0000000E+00 | -7.6246471E-02 | -3.9143378E-02 | -2.4355044E-02 | -1.8027734E-02 |
| 2.0E-01 | 0.0000000E+00 | -5.5002858E-02 | -4.2111775E-02 | -2.6285697E-02 | -1.8499005E-02 |
| 3.0E-01 | 0.0000000E+00 | -4.0385928E-02 | -4.2154954E-02 | -2.7603362E-02 | -1.9064225E-02 |
| 4.0E-01 | 0.0000000E+00 | -3.0241486E-02 | -3.9182918E-02 | -2.7410799E-02 | -1.9177355E-02 |
| 5.0E-01 | 0.0000000E+00 | -2.2667335E-02 | -3.4237609E-02 | -2.5515677E-02 | -1.8383931E-02 |
| 6.0E-01 | 0.0000000E+00 | -1.6652872E-02 | -2.8158045E-02 | -2.2160613E-02 | -1.6519226E-02 |
| 7.0E-01 | 0.0000000E+00 | -1.1644990E-02 | -2.1457784E-02 | -1.7675673E-02 | -1.3619699E-02 |
| 8.0E-01 | 0.0000000E+00 | -7.3213018E-03 | -1.4431859E-02 | -1.2349066E-02 | -9.8072984E-03 |
| 9.0E-01 | 0.0000000E+00 | -3.4826882E-03 | -7.2476382E-03 | -6.4027104E-03 | -5.2233471E-03 |
| 1.0E+00 | 0.0000000E+00 | 0.0000000E+00 | 0.0000000E+00 | 0.0000000E+00 | 0.0000000E+00 |

## Table 8a. Component $I(\eta\tau, \mu)$ Conservative Medium.

| $\mu \backslash \eta$ | 0.0 | 0.1 | 0.5 | 0.75 | 1.0 |
|---|---|---|---|---|---|
| -1.0E+00 | 5.3857832E-01 | 5.1525547E-01 | 3.4560447E-01 | 2.2233516E-01 | 1.1008408E-01 |
| -9.0E-01 | 5.4598454E-01 | 5.2606667E-01 | 3.5844560E-01 | 2.3042795E-01 | 1.1008408E-01 |
| -8.0E-01 | 5.5380538E-01 | 5.3785743E-01 | 3.7328745E-01 | 2.4018399E-01 | 1.1008408E-01 |
| -7.0E-01 | 5.6168113E-01 | 5.5043239E-01 | 3.9045050E-01 | 2.5208478E-01 | 1.1008408E-01 |
| -6.0E-01 | 5.6901960E-01 | 5.6340321E-01 | 4.1025390E-01 | 2.6679888E-01 | 1.1008408E-01 |
| -5.0E-01 | 5.7489445E-01 | 5.7608671E-01 | 4.3292242E-01 | 2.8525948E-01 | 1.1008408E-01 |
| -4.0E-01 | 5.7793895E-01 | 5.8739766E-01 | 4.5835721E-01 | 3.0873773E-01 | 1.1008408E-01 |
| -3.0E-01 | 5.7628018E-01 | 5.9583025E-01 | 4.8564074E-01 | 3.3874672E-01 | 1.1008408E-01 |
| -2.0E-01 | 5.6744156E-01 | 5.9968757E-01 | 5.1232375E-01 | 3.7591828E-01 | 1.1008408E-01 |
| -1.0E-01 | 5.4712744E-01 | 5.9717767E-01 | 5.3525255E-01 | 4.1462116E-01 | 1.1008408E-01 |
| 0.0E+00 | 4.9487896E-01 | 5.8451246E-01 | 5.5452334E-01 | 4.4342942E-01 | 1.1008408E-01 |
| 0.0E+00 | 0.0000000E+00 | 5.8451246E-01 | 5.5452334E-01 | 4.4342942E-01 | 2.7910661E-01 |
| 1.0E-01 | 0.0000000E+00 | 4.8836921E-01 | 5.7002524E-01 | 4.6838343E-01 | 3.2007662E-01 |
| 2.0E-01 | 0.0000000E+00 | 3.5519545E-01 | 5.7631025E-01 | 4.8987460E-01 | 3.5185482E-01 |
| 3.0E-01 | 0.0000000E+00 | 2.7464291E-01 | 5.6572066E-01 | 5.0426151E-01 | 3.7893524E-01 |
| 4.0E-01 | 0.0000000E+00 | 2.2421285E-01 | 5.4353027E-01 | 5.0919019E-01 | 4.0000895E-01 |
| 5.0E-01 | 0.0000000E+00 | 1.9039923E-01 | 5.1705540E-01 | 5.0631937E-01 | 4.1448624E-01 |
| 6.0E-01 | 0.0000000E+00 | 1.6648094E-01 | 4.9048463E-01 | 4.9846030E-01 | 4.2316139E-01 |
| 7.0E-01 | 0.0000000E+00 | 1.4887992E-01 | 4.6570019E-01 | 4.8793132E-01 | 4.2738001E-01 |
| 8.0E-01 | 0.0000000E+00 | 1.3554256E-01 | 4.4338753E-01 | 4.7629934E-01 | 4.2845085E-01 |
| 9.0E-01 | 0.0000000E+00 | 1.2521294E-01 | 4.2367194E-01 | 4.6453504E-01 | 4.2743733E-01 |
| 1.0E+00 | 0.0000000E+00 | 1.1708207E-01 | 4.0643463E-01 | 4.5320674E-01 | 4.2513624E-01 |

## Table 8b. Component $Q(\eta\tau, \mu)$ Conservative Medium.

| $\mu \backslash \eta$ | 0.0 | 0.1 | 0.5 | 0.75 | 1.0 |
|---|---|---|---|---|---|
| -1.0E+00 | 0.0000000E+00 | 0.0000000E+00 | 0.0000000E+00 | 0.0000000E+00 | 0.0000000E+00 |
| -9.0E-01 | -9.8625348E-03 | -7.8913743E-03 | -3.2138937E-03 | -1.5725107E-03 | 0.0000000E+00 |
| -8.0E-01 | -1.9810574E-02 | -1.5861721E-02 | -6.5291172E-03 | -3.2527646E-03 | 0.0000000E+00 |
| -7.0E-01 | -2.9845898E-02 | -2.3910027E-02 | -9.9572185E-03 | -5.0698510E-03 | 0.0000000E+00 |
| -6.0E-01 | -3.9968813E-02 | -3.2030920E-02 | -1.3505924E-02 | -7.0626198E-03 | 0.0000000E+00 |
| -5.0E-01 | -5.0180316E-02 | -4.0214259E-02 | -1.7171816E-02 | -9.2818815E-03 | 0.0000000E+00 |
| -4.0E-01 | -6.0489907E-02 | -4.8448978E-02 | -2.0925170E-02 | -1.1788256E-02 | 0.0000000E+00 |
| -3.0E-01 | -7.0935479E-02 | -5.6738852E-02 | -2.4684068E-02 | -1.4627841E-02 | 0.0000000E+00 |
| -2.0E-01 | -8.1618567E-02 | -6.5136423E-02 | -2.8302168E-02 | -1.7711540E-02 | 0.0000000E+00 |
| -1.0E-01 | -9.2752093E-02 | -7.3760995E-02 | -3.1722361E-02 | -2.0403287E-02 | 0.0000000E+00 |
| 0.0E+00 | -1.0532884E-01 | -8.2787161E-02 | -3.5196046E-02 | -2.2164172E-02 | 0.0000000E+00 |
| 0.0E+00 | 0.0000000E+00 | -8.2787161E-02 | -3.5196046E-02 | -2.2164172E-02 | -1.8620295E-02 |
| 1.0E-01 | 0.0000000E+00 | -7.6851215E-02 | -3.8835042E-02 | -2.4010691E-02 | -1.8070261E-02 |
| 2.0E-01 | 0.0000000E+00 | -5.5459385E-02 | -4.1884194E-02 | -2.5952301E-02 | -1.8447531E-02 |
| 3.0E-01 | 0.0000000E+00 | -4.0726475E-02 | -4.2006406E-02 | -2.7309217E-02 | -1.8964917E-02 |
| 4.0E-01 | 0.0000000E+00 | -3.0498526E-02 | -3.9092818E-02 | -2.7167952E-02 | -1.9062057E-02 |
| 5.0E-01 | 0.0000000E+00 | -2.2860917E-02 | -3.4186659E-02 | -2.5324807E-02 | -1.8272604E-02 |
| 6.0E-01 | 0.0000000E+00 | -1.6795542E-02 | -2.8132033E-02 | -2.2017709E-02 | -1.6423205E-02 |
| 7.0E-01 | 0.0000000E+00 | -1.1744982E-02 | -2.1446782E-02 | -1.7575498E-02 | -1.3544952E-02 |
| 8.0E-01 | 0.0000000E+00 | -7.3842743E-03 | -1.4428950E-02 | -1.2286539E-02 | -9.7566779E-03 |
| 9.0E-01 | 0.0000000E+00 | -3.5126832E-03 | -7.2479381E-03 | -6.3733500E-03 | -5.1979609E-03 |
| 1.0E+00 | 0.0000000E+00 | 0.0000000E+00 | 0.0000000E+00 | 0.0000000E+00 | 0.0000000E+00 |

**C Randomness**

As mentioned, the adding and doubling method naturally accommodates heterogeneous media. To test this, we assemble a random distribution of edits for C2/Case 1 for the homogeneous medium and compare the exiting distribution to the established benchmark of Tables 7. Since the medium is still homogeneous, the distributions should be identical. In addition, we apply a new computational strategy, which is to simply set $l$ and $N$ to 5 and 24 respectively and forgo acceleration in lieu of having Tables 7. When the medium is composed of 75 random intervals, the identical exiting intensity distribution of Tables 7 results further providing confidence in the robustness of the method.

A second demonstration assumes a uniform distribution of edits for the medium of C2/Case 1, but with random absorption ($\omega$) in the three nodes on either side of center ($\tau=1$). Otherwise, the medium is the same as C2/Case 1 as shown in Table 9. This is a simulation of polarization in damaged material.

**Table 9. Damaged Medium.**

| Node(j) | $\omega$ | $\tau_j$ | c |
|---|---|---|---|
| 31 | ……. | …… | ….. |
| 32 | 0.9900000000000000 | 0.8533333333333334 | 0.500000 |
| 33 | 0.9900000000000000 | 0.8800000000000000 | 0.500000 |
| 34 | 0.7399273383434252 | 0.9066666666666667 | 0.500000 |
| 35 | 0.7042137833434735 | 0.9333333333333333 | 0.500000 |
| 36 | 0.8654406633936658 | 0.9600000000000001 | 0.500000 |
| 37 | 0.3732005428851675 | 0.9866666666666667 | 0.500000 |
| 38 | 5.972673098081264E-02 | 1.013333333333333 | 0.500000 |
| 39 | 0.8133076113680893 | 1.040000000000000 | 0.500000 |
| 40 | 0.2310978585941060 | 1.066666666666667 | 0.500000 |
| 41 | 0.9900000000000000 | 1.093333333333333 | 0.500000 |
| 42 | 0.9900000000000000 | 1.120000000000000 | 0.500000 |
| 43 | ….. | ….. | ….. |

The two top plates in Fig.4 are the intensity $I$ and $Q$ contours for the undamaged medium and the two bottom plates are the contours for the damaged medium. A blue band highlights the damaged region . While it is difficult to visually interpret the contour plots, one can at least say that for $I$, the effect of the damage is considerably more than for $Q$. The intensity in the first half of the medium is unaffected by the damage; while, there is a concentration of the effect at the slab center and grazing direction for both components. This exercise is an example of the kind of investigation that is possible through doubling.

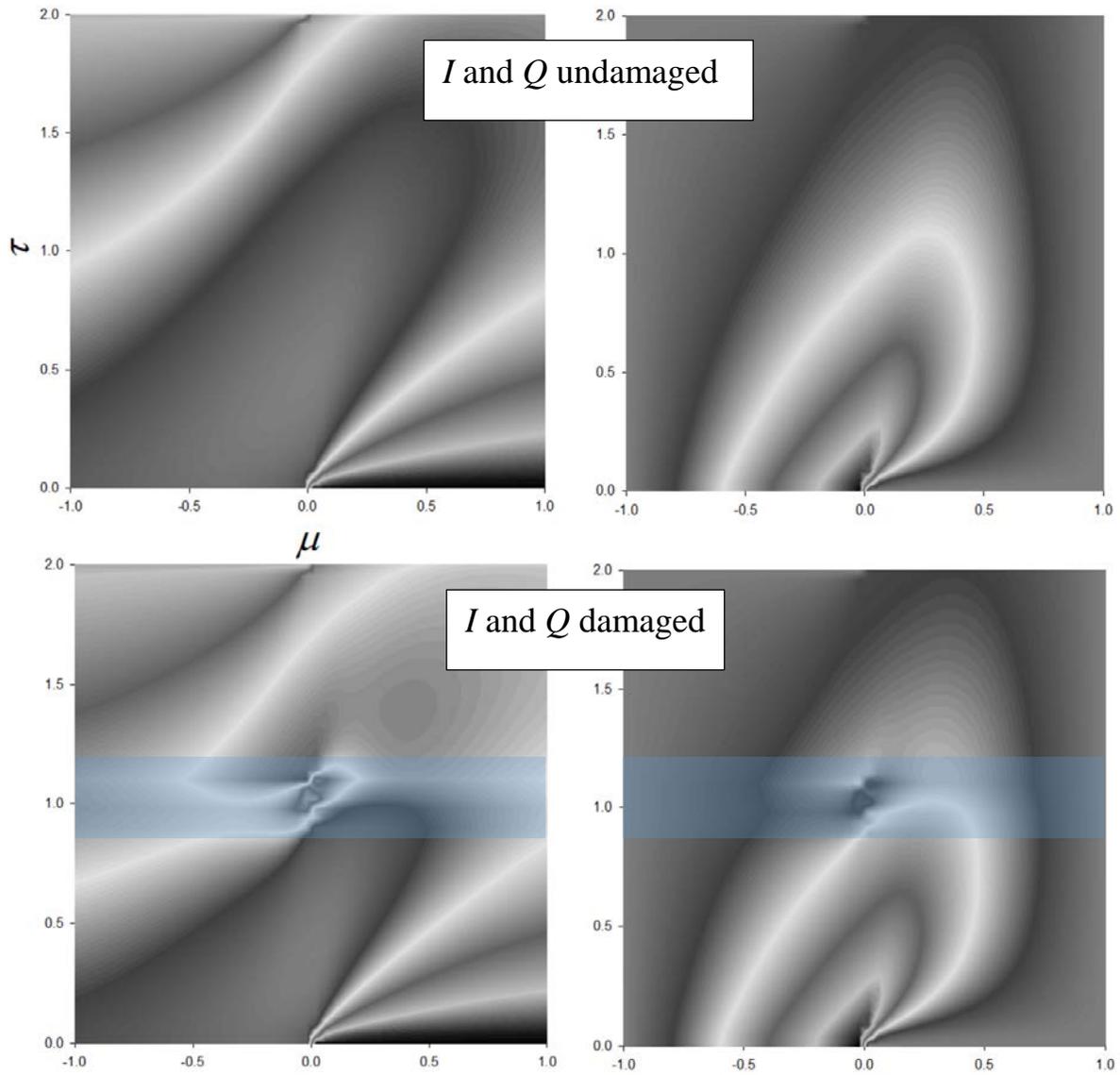

Fig. 4. Intensity contours from undamaged and damaged polarizing media.

In closing the results section, the exiting distribution for the damaged medium is presented in Table 9 as a benchmark. The entries should be precise to one unit in the last quoted digit and required about a minute of computational time.

**Table 9. Exiting Distribution for Damaged Medium.**

| $\mu$ | $I(0,\mu)$ | $I(\tau_0,\mu)$ | $Q(0,\mu)$ | $Q(\tau_0,\mu)$ |
|---|---|---|---|---|
| -1.0000000E+00 | 4.3990337E-01 | 9.0799486E-02 | 0.0000000E+00 | 0.0000000E+00 |
| -9.0000000E-01 | 4.4604915E-01 | 9.0799486E-02 | -9.8323910E-03 | 0.0000000E+00 |
| -8.0000000E-01 | 4.5313813E-01 | 9.0799486E-02 | -1.9746460E-02 | 0.0000000E+00 |
| -7.0000000E-01 | 4.6110044E-01 | 9.0799486E-02 | -2.9742867E-02 | 0.0000000E+00 |
| -6.0000000E-01 | 4.6975972E-01 | 9.0799486E-02 | -3.9819930E-02 | 0.0000000E+00 |
| -5.0000000E-01 | 4.7875939E-01 | 9.0799486E-02 | -4.9974414E-02 | 0.0000000E+00 |
| -4.0000000E-01 | 4.8742087E-01 | 9.0799486E-02 | -6.0205214E-02 | 0.0000000E+00 |
| -3.0000000E-01 | 4.9439221E-01 | 9.0799486E-02 | -7.0520546E-02 | 0.0000000E+00 |
| -2.0000000E-01 | 4.9659685E-01 | 9.0799486E-02 | -8.0940354E-02 | 0.0000000E+00 |
| -1.0000000E-01 | 4.8714105E-01 | 9.0799486E-02 | -9.1538723E-02 | 0.0000000E+00 |
| 0.0000000E+00 | 4.4700710E-01 | 9.0799486E-02 | -1.0316880E-01 | 0.0000000E+00 |
| 0.0000000E+00 | 0.0000000E+00 | 2.2262605E-01 | 0.0000000E+00 | -1.7637547E-02 |
| 1.0000000E-01 | 0.0000000E+00 | 2.5377831E-01 | 0.0000000E+00 | -1.7781076E-02 |
| 2.0000000E-01 | 0.0000000E+00 | 2.7620237E-01 | 0.0000000E+00 | -1.8521399E-02 |
| 3.0000000E-01 | 0.0000000E+00 | 2.9341681E-01 | 0.0000000E+00 | -1.9162511E-02 |
| 4.0000000E-01 | 0.0000000E+00 | 3.0678254E-01 | 0.0000000E+00 | -1.9264955E-02 |
| 5.0000000E-01 | 0.0000000E+00 | 3.1645184E-01 | 0.0000000E+00 | -1.8442421E-02 |
| 6.0000000E-01 | 0.0000000E+00 | 3.2277300E-01 | 0.0000000E+00 | -1.6551775E-02 |
| 7.0000000E-01 | 0.0000000E+00 | 3.2642629E-01 | 0.0000000E+00 | -1.3634129E-02 |
| 8.0000000E-01 | 0.0000000E+00 | 3.2815172E-01 | 0.0000000E+00 | -9.8111711E-03 |
| 9.0000000E-01 | 0.0000000E+00 | 3.2859158E-01 | 0.0000000E+00 | -5.2228676E-03 |
| 1.0000000E+00 | 0.0000000E+00 | 3.2824740E-01 | 0.0000000E+00 | 0.0000000E+00 |

## CONCLUSION

Once again, the method of doubling and adding has been applied to solve the transport equation. Though the method is one of the oldest, it is also one of the most precise and conceptually straightforward. It is limited however to media without volume sources unless a particular solution is handy. In this work, adding and doubling is applied to the case of a polarizing slab medium in the Stokes 2-vector approximation. The novelty here is inclusion of convergence acceleration through Wynn-epsilon and Richardsons algorithms to enhance the precision of the solution to achieve the stated goal of generating highly precise benchmarks. To achieve a high precision benchmark, special care must be exercised in the numerical implementation of adding and doubling. In particular, a new numerically equivalent representation of the delta function enabled the method without recourse to a particular solution. The grazing direction was treated by simply setting the first (derivative) term of the transport equation to zero and populating the RHS with the known intensities. A "faux quadrature" allowed a consistent relative error across quadrature orders to monitor convergence of the sequence of solutions generated by decreasing the spatial discretization and increasing the quadrature order. These are the sequences that are accelerated to convergence.

Four highly precise benchmarks were generated. Two improved upon already existing benchmarks. A third modified an existing benchmark to treat the conservative case and a fourth showed how doubling accommodates random heterogeneous media.

The takeaway message of this presentation is that a benchmark solution, in order to be a true benchmark, should have an internal mechanism to verify the numerical method used. Convergence acceleration is such a mechanism and its value has been shown here.

**Appendix A: Explicit Forms for $A^n$**
When the matrix $A$ of Eq(5c) is

$$A = \begin{bmatrix} M^{-1}(I-PW) & -M^{-1}PW \\ M^{-1}PW & -M^{-1}(I-PW) \end{bmatrix} \equiv \begin{bmatrix} \alpha & -\beta \\ \beta & -\alpha \end{bmatrix} = \begin{bmatrix} \eta_{11}^{(1)} & -\eta_{12}^{(1)} \\ \eta_{12}^{(1)} & -\eta_{11}^{(1)} \end{bmatrix}, \quad (A1)$$

we find for $n = 2$

$$A^2 = \begin{bmatrix} \alpha & -\beta \\ \beta & -\alpha \end{bmatrix} \begin{bmatrix} \alpha & -\beta \\ \beta & -\alpha \end{bmatrix} = \begin{bmatrix} \alpha^2 - \beta^2 & -\alpha\beta + \alpha\beta \\ \alpha\beta - \alpha\beta & -\beta^2 + \alpha^2 \end{bmatrix} = \begin{bmatrix} \gamma_{11}^{(1)} & \gamma_{12}^{(1)} \\ \gamma_{12}^{(1)} & \gamma_{11}^{(1)} \end{bmatrix}. \quad (A2)$$

Assume the conjecture that $A$ to the even power ($n = 2m$) is symmetric

$$A^{2m} = \begin{bmatrix} \gamma_{11}^{(m)} & \gamma_{12}^{(m)} \\ \gamma_{12}^{(m)} & \gamma_{11}^{(m)} \end{bmatrix}, \quad (A3)$$

which follows inductively by construction

$$\begin{aligned}
\boldsymbol{A}^{2m} = \boldsymbol{A}^{2(m-1)}\boldsymbol{A}^2 &= \begin{bmatrix} \gamma_{11}^{(m-1)} & \gamma_{12}^{(m-1)} \\ \gamma_{12}^{(m-1)} & \gamma_{11}^{(m-1)} \end{bmatrix} \begin{bmatrix} \gamma_{11}^{(1)} & \gamma_{12}^{(1)} \\ \gamma_{12}^{(1)} & \gamma_{11}^{(1)} \end{bmatrix} \\
&= \begin{bmatrix} \gamma_{11}^{(m-1)}\gamma_{11}^{(1)} + \gamma_{12}^{(m-1)}\gamma_{12}^{(1)} & \gamma_{11}^{(m-1)}\gamma_{12}^{(1)} + \gamma_{12}^{(m-1)}\gamma_{11}^{(1)} \\ \gamma_{12}^{(m-1)}\gamma_{11}^{(1)} + \gamma_{11}^{(m-1)}\gamma_{12}^{(1)} & \gamma_{12}^{(m-1)}\gamma_{12}^{(1)} + \gamma_{11}^{(m-1)}\gamma_{11}^{(1)} \end{bmatrix} \\
&= \begin{bmatrix} \gamma_{11}^{(m)} & \gamma_{12}^{(m)} \\ \gamma_{12}^{(m)} & \gamma_{11}^{(m)} \end{bmatrix}.
\end{aligned} \qquad (A4)$$

For $\boldsymbol{A}$ to the odd power ($n = 2m-1$), we again show inductively

$$\begin{aligned}
\boldsymbol{A}^{2m-1} = \boldsymbol{A}^{2(m-1)}\boldsymbol{A} &= \begin{bmatrix} \gamma_{11}^{(m-1)} & \gamma_{12}^{(m-1)} \\ \gamma_{12}^{(m-1)} & \gamma_{11}^{(m-1)} \end{bmatrix} \begin{bmatrix} \eta_{11}^{(1)} & -\eta_{12}^{(1)} \\ \eta_{12}^{(1)} & -\eta_{11}^{(1)} \end{bmatrix} \\
&= \begin{bmatrix} \gamma_{11}^{(m-1)}\eta_{11}^{(1)} + \gamma_{12}^{(m-1)}\eta_{12}^{(1)} & -\gamma_{11}^{(m-1)}\eta_{12}^{(1)} - \gamma_{12}^{(m-1)}\eta_{11}^{(1)} \\ \gamma_{12}^{(m-1)}\eta_{11}^{(1)} + \gamma_{11}^{(m-1)}\eta_{12}^{(1)} & -\gamma_{12}^{(m-1)}\eta_{12}^{(1)} - \gamma_{11}^{(m-1)}\eta_{11}^{(1)} \end{bmatrix} \\
&= \begin{bmatrix} \eta_{11}^{(m)} & -\eta_{12}^{(m)} \\ \eta_{12}^{(m)} & -\eta_{11}^{(m)} \end{bmatrix}.
\end{aligned} \qquad (A5)$$

indicating the sign pattern is invariant. When inserted into the Taylor series or polynomial for even and odd powers of $\boldsymbol{A}$, Eqs(17) and (19) result.

## Appendix B: Sufficient Condition for a Symmetric Response Matrix

To determine a sufficient condition for a symmetric response matrix, we begin with the partitioned response matrix

$$\boldsymbol{R} = \begin{bmatrix} -\boldsymbol{P}_{12} & \boldsymbol{P}_{11}^* \\ -\boldsymbol{P}_{22} & \boldsymbol{P}_{21}^* \end{bmatrix}^{-1} \begin{bmatrix} -\boldsymbol{P}_{12}^* & \boldsymbol{P}_{11} \\ -\boldsymbol{P}_{22}^* & \boldsymbol{P}_{21} \end{bmatrix}. \qquad (B1)$$

Assume that $\boldsymbol{P}$ satisfies

$$\boldsymbol{P} = \begin{bmatrix} \boldsymbol{P}_{22}^* & \boldsymbol{P}_{21}^* \\ \boldsymbol{P}_{12}^* & \boldsymbol{P}_{11}^* \end{bmatrix}, \qquad (B2b)$$

which is true when $P(h) = P^*(-h)$. This assumption gives Eq(45b)

$$R \equiv \begin{bmatrix} -P_{21}^* & P_{11}^* \\ -P_{11}^* & P_{21}^* \end{bmatrix}^{-1} \begin{bmatrix} -P_{12}^* & P_{22}^* \\ -P_{22}^* & P_{12}^* \end{bmatrix}. \tag{B3}$$

for the response matrix. If one multiplies the bottom row by $-1$ and exchanges it with the top row, then

$$R \equiv \begin{bmatrix} P_{11}^* & -P_{21}^* \\ -P_{21}^* & P_{11}^* \end{bmatrix}^{-1} \begin{bmatrix} P_{22}^* & -P_{12}^* \\ -P_{12}^* & P_{22}^* \end{bmatrix}. \tag{B4}$$

An explicit expression for the inverse comes from

$$T^* \equiv \begin{bmatrix} P_{11}^* & -P_{21}^* \\ -P_{21}^* & P_{11}^* \end{bmatrix} = \begin{bmatrix} T_1^* & T_2^* \\ T_3^* & T_4^* \end{bmatrix}, \tag{B5}$$

which we solve for the partitions through

$$TT^* \equiv \begin{bmatrix} I & 0 \\ 0 & I \end{bmatrix} \tag{B6}$$

to give (after some algebra)

$$T_1^* = T_4^* = \frac{1}{2}\left[\left(P_{11}^* - P_{21}^*\right)^{-1} + \left(P_{11}^* + P_{21}^*\right)^{-1}\right]$$

$$T_2^* = T_3^* = \frac{1}{2}\left[\left(P_{11}^* - P_{21}^*\right)^{-1} - \left(P_{11}^* + P_{21}^*\right)^{-1}\right]. \tag{B7a,b}$$

The response matrix is therefore

$$R \equiv \begin{bmatrix} T_1^* & T_2^* \\ T_3^* & T_4^* \end{bmatrix} \begin{bmatrix} P_{22}^* & -P_{12}^* \\ -P_{12}^* & P_{22}^* \end{bmatrix} = \begin{bmatrix} T_n & R_f \\ R_f & T_n \end{bmatrix}. \tag{B8}$$

From Eq(B8), on matrix multiplication, the following symmetry:

$$\begin{bmatrix} \boldsymbol{T}_n \\ \boldsymbol{R}_f \end{bmatrix} = \frac{1}{2}\left[ \left(\boldsymbol{P}_{11}^* - \boldsymbol{P}_{21}^*\right)^{-1}\left(\boldsymbol{P}_{22}^* - \boldsymbol{P}_{12}^*\right) \pm \left(\boldsymbol{P}_{11}^* + \boldsymbol{P}_{21}^*\right)^{-1}\left(\boldsymbol{P}_{22}^* + \boldsymbol{P}_{12}^*\right) \right] \qquad (B9)$$

emerges.